\documentclass[12pt,a4paper]{article}  
\usepackage{hyperref, amsmath, amssymb, graphicx, amsfonts, latexsym, bbold, mathbbol}
\usepackage{slashed}
\usepackage[]{hyperref}

\usepackage{tikz}
\usepackage{tikz-feynman}
\tikzfeynmanset{compat=1.0.0}

\newcommand{\be}{\begin{equation}}
\newcommand{\ee}{\end{equation}}
\newcommand{\ba}{\begin{equation} \begin{aligned}}
\newcommand{\ea}{\end{aligned} \end{equation}}

\begin{document}

\begin{center}

{\LARGE \bf  Chiral fermions, dimensional regularization, \\[2mm] and the trace anomaly}

\vskip 1.2cm
Fiorenzo Bastianelli$^{\,a,b}$  and Luca Chiese$^{\,a}$ 
\vskip 1cm

$^a${\em Dipartimento di Fisica e Astronomia ``Augusto Righi", Universit{\`a} di Bologna,\\
via Irnerio 46, I-40126 Bologna, Italy}

$^b${\em   INFN, Sezione di Bologna, via Irnerio 46, I-40126 Bologna, Italy}
 
\end{center}
\vskip .8cm

\abstract{We investigate the trace anomaly of a chiral fermion in dimensional regularization, 
considering in detail the simplest case of coupling to an abelian gauge field.
We apply the Breitenlohner-Maison/'t Hooft-Veltman prescription for dealing with the chiral matrix 
$\gamma^5$ and verify that no parity-odd term arises in the trace anomaly.
The issue treated here is analogous to that of a chiral fermion in curved spacetime,
discussed in recent literature. The advantage of having a simplified background minimizes 
the amount of algebraic calculations and allows pinpointing better
the subtle points carried by dimensional regularization.}

\section{Introduction}
The structure of the trace anomaly of a chiral fermion has been the focus of a recent debate. 
The main issue concerns the presence or absence of a parity-odd term in the trace anomaly 
of a chiral fermion in a curved background. While such a term was conjectured to be a possibility
in certain chiral theories \cite{Nakayama:2012gu}, it is generically unexpected for the case 
of a Weyl fermion. However, an explicit calculation performed in dimensional regularization 
seemed to indicate its presence \cite{Bonora:2014qla}. 
Several groups have been unable to verify that claim, either directly or indirectly,
\cite{Bastianelli:2016nuf, Godazgar:2018boc, Frob:2019dgf, Bastianelli:2019zrq, Nakagawa:2020gqc, Abdallah:2021eii},
and so far the only confirmations  \cite{Bonora:2015nqa, Bonora:2017gzz, Bonora:2018obr}
are from groups related to Ref. \cite{Bonora:2014qla}.

We believe that the arguments and proofs on the absence of parity-odd terms 
in the trace anomaly, discussed in the above papers and references therein, are sufficiently solid. However, we find it important to clarify as much as possible the calculational 
details in support of general arguments, as to debug them 
(in one sense or the other) from errors.
This is particularly desirable regarding the use of dimensional regularization. 
Dimensional regularization is notoriously subtle in chiral theories.
In the case of the trace anomaly of a Weyl fermion
it has been used in \cite{Bonora:2014qla} and \cite{Abdallah:2021eii},
with the latter presenting a critical assessment of the former,
see also \cite{Abdallah:2022okt}.
Here we wish to offer a similar application of dimensional regularization
in the simplified context of a chiral fermion coupled to an abelian gauge field.
The algebraic structure of the Feynman graphs is simpler
compared to the one arising from the coupling to gravity,
and all terms entering the trace anomaly can be computed by hand.
The strategy of employing a simplified background
has already been considered in \cite{Bastianelli:2018osv}
(and \cite{Bastianelli:2019fot} for the non-abelian case),
where Pauli-Villars regularization and heat kernel techniques 
were used to compute the trace anomaly for a Weyl fermion coupled to a gauge field, 
thus proving the absence of parity-odd terms in that case. 
Here we wish to apply dimensional regularization to the same problem,
and check that result. We report explicitly the technical details of the calculation 
in dimensional regularization, as they may contribute to dispel 
doubts---present in the scientific community, see e.g. Ref. \cite{Chalabi:2021jud}---suggesting 
that the issue is still unsettled.  

One reason why a CP odd contribution to the trace anomaly cannot appear is 
related to the CP invariance of the Weyl fermion lagrangian.
If the regularization adopted could manifestly preserve this symmetry, then the absence of CP odd terms
in the trace anomaly would be assured. This is the case discussed in ref.  \cite{Bastianelli:2018osv}, were
a regularization with  Pauli-Villars fields with Majorana mass could manifestly preserve CP invariance. 
However, dimensional regularization 
does not preserve chiral properties, and the result is not guaranteed a priori.
However, as we shall see, the use of the charge conjugation matrix $C$ will help in proving 
the cancellation of the  CP odd terms in dimensional regularization.

In the following, we start by briefly reviewing the lagrangian of a Weyl fermion coupled 
to an abelian gauge field and present the corresponding classical conservation laws. 
Then, we turn to the perturbative calculations of the anomalies.
We first review the well-known case of the chiral anomaly to set the stage, 
and then turn to the calculation of the trace anomaly, showing 
that no parity-odd term may arise. 
Finally, we check the Ward identities related to the conservation of the stress tensor, 
to make sure that no gravitational anomaly is induced by dimensional regularization. 
Our conventions follow those of Ref. \cite{Bastianelli:2018osv} and are 
recapitulated in appendix \ref{A}. 
We report the main points of the Breitenlohner-Maison-'t Hooft-Veltman prescription
in appendix \ref{B}, while computational details are collected in appendices \ref{C} and \ref{D}.

\section{Lagrangian and classical conservations laws} 

A Weyl fermion $\lambda(x)$ coupled to an abelian gauge field $A_a(x)$ is
described by the lagrangian 
\begin{equation}
\mathcal{L}= - \bar{\lambda}\slashed{D}\lambda 
\label{lag}
\end{equation}
with $ \slashed{D}= \gamma^a D_a$, where $ D_a = \partial_a - i A_a $ is the gauge covariant derivative acting on $\lambda$. 
We have taken
$\lambda$ to be left-handed, $\lambda=P_L \lambda$, while its Dirac conjugate satisfies 
$\bar{\lambda}= \bar{\lambda}P_R$, with chiral projectors defined by 
 \begin{equation} \label{chiral projectors}
 P_L= \frac{1}{2}\left(\mathbb{1} + \gamma^5 \right) \;, \qquad P_R =\frac{1}{2}\left(\mathbb{1} - \gamma^5 \right) \;.
 \end{equation}
 Chirality forbids a Dirac mass, while a Majorana mass is incompatible with the gauge symmetry.
 
 The gauge current that couples to the abelian gauge field $A_a$ is given by
 \begin{equation} 
  J^a= i \bar{\lambda} \gamma^a \lambda
\label{chiral current}
\end{equation}
and is conserved on-shell because of the gauge symmetry,
 \begin{equation} 
\partial_a J^a=0 \;.
\label{c1}
\end{equation}
It develops a well-known gauge anomaly at the quantum level.

The stress tensor can be obtained by coupling the theory to gravity through a vielbein 
$e_\mu{}^a$, 
and defined by the functional derivative of the action $S$ with respect to the vielbein,  
$T^{\mu}{}_ a=\frac{1}{e} \frac{\delta S}{\delta e_\mu{}^a}$.
In the flat space limit, dropping terms that vanish upon the use of the equations of motion, 
it takes the form
 \begin{equation} \label{stress tensor}
 T^{ab}=\frac{1}{4} \bar{\lambda} \big ( 
 \gamma^a \overset{\leftrightarrow}{D^b} + \gamma^b \overset{\leftrightarrow}{D^a} \big ) \lambda 
 \end{equation}
 where $\overset{\leftrightarrow}{D_a} = D_a - \overset{\leftarrow}{D_a} $
 and $\overset{\leftarrow}{D_a}= \overset{\leftarrow}{\partial_a} + i A_a $. 
 It is manifestly symmetric and satisfies the equations
\begin{align}
\partial_a  T^{ab} &=  - J_a F^{ab} \label{c2}\\
T^a_{\ a} &  =0     \label{c3}
 \end{align}        
which are valid on-shell.
The first one is due to translational invariance, broken classically by the background field 
$A_a$ with field strength $F_{ab} =\partial_a A_b -\partial_b A_a$.
It does not develop any quantum anomaly.
The second relation follows from the Weyl invariance in curved space and is related to conformal invariance 
in flat space. It develops a trace anomaly. In the following, we want to make sure
that this trace anomaly does not contain any parity-odd contribution.
Of course, eqs. \eqref{c1}, \eqref{c2}, \eqref{c3} may be verified directly using the 
classical equations of motion of $\lambda$ and $\bar{\lambda}$. 
 
Let us close this section with a few comments on CP invariance, which help in interpreting 
 our results on the trace anomaly, as mentioned in the introduction.
The Weyl theory breaks parity, as the parity transformation of a generic Dirac spinor $\psi$
\be
\psi(x)\quad 
\stackrel{P}{\longrightarrow} \quad 
\psi_{_P}(x_{_P}) = \eta_{_P} \beta \psi(x)
\ee
is incompatible with the chiral properties of the Weyl spinor $\lambda$ (with the constraint $ \lambda =\gamma^5\lambda$).
Here $x_{_P}=(t, -\vec{x})$, $\eta_{_P}$ is a phase that can be chosen to be either $\pm1$ or $\pm i$,
and $\beta=i\gamma^0$ is the matrix that sits in the definition of the Dirac
conjugate spinor ($\bar{\psi}=\psi^\dagger \beta$, we follow the notations spelled out in appendix A of  \cite{Bastianelli:2018osv}).
However, one can combine it with charge conjugation, that acts on a generic Dirac spinor as
\be
\psi(x) \quad \stackrel{C}{\longrightarrow}
\quad   \psi_{_C}(x) =  \eta_{_C} C \bar{\psi}^T(x)
\ee
and that by itself would again be incompatible with the chirality of the Weyl spinor $\lambda$.
The charge conjugation matrix $C$ satisfies $C \gamma^a C^{-1} = - \gamma^{a\, T}$,
while $\eta_{_C}$ is an arbitrary phase.
The combination of C and P gives rise to the CP transformation, that is now compatible with the chiral properties of a Weyl spinor 
and reads
\be
\lambda(x) \quad \stackrel{CP}{\longrightarrow}
\quad   \lambda_{_{CP}}(x_{_P}) = 
 \eta_{_{CP}} C \lambda^* (x) \;.
\ee
The phase $\eta_{_{CP}}$  is unobservable for a single fermion and could be set to one.
Then, it is easy to check that the action from the lagrangian \eqref{lag} 
is invariant (integrals of total derivatives are discarded and the jacobian from changing 
variables from $x_{_P}$ to $x$ is unity) if the background gauge field is also CP transformed as usual.
 
\section{The chiral anomaly} 

As a preparation, we start repeating the exercise of computing the chiral anomaly. 
Though standard---it is material that may be found on various QFT textbooks--- it 
sets the stage for our subsequent calculations. 
We regularize the Feynman diagrams  with dimensional regularization, extended by
the Breitenlohner-Maison/'t Hooft-Veltman prescription for  dealing with the chiral 
matrix $\gamma^5$, see appendix \ref{B}.

In order to derive the chiral anomaly, we expand in perturbation theory
the  expectation value $\langle J^c (x)\rangle_{_{\!A}}$
of the axial current $J^c (x) = i \bar \lambda (x) \gamma^c \lambda (x)$ 
in the background of the abelian gauge field $A_a$, and 
check the Ward identity that follows from eq. \eqref{c1}.
Perturbation theory  is generated by setting 
$\langle J^c (x)\rangle_{_{\!A}} = \langle J^c (x)\,  e^{i S_{int}}\rangle$, 
with $S_{int}=  \int d^4x\, A_a(x) J^a(x) $ and using the perturbative propagator 
\begin{equation} 
\langle \lambda(x)  \bar \lambda (y) \rangle =  
\int {d^4p\over (2\pi)^4}\,   e^{ip (x-y)}   P_L \frac{-\slashed{p}}{p^2} P_R 
\end{equation}
where we have kept explicitly both projectors matching the chiral properties of the left-hand side
and denoted $px\equiv p_a x^a$.

The chiral anomaly arises at second order in the gauge field
\begin{align} 
\langle J^c(x) \rangle_{_{\!A}}^{^{\! (2)}}   
&= - \frac{1}{2} \int d^4 y \int d^4 z \, A_a(y) A_b(z) \,  \langle J^c(x) J^a (y) J^b(z) \rangle 
\nonumber 
\\
&= - \frac{1}{2} \int d^4 y \int d^4 z\, A_a(y) A_b(z)
		\int \frac{d^4 k}{(2 \pi)^4}  \int \frac{d^4 p}{(2 \pi)^4} \int \frac{d^4 q}{(2 \pi)^4}  \nonumber
		 \\
	& \quad \quad  \times e^{-ikx} e^{ipy} e ^{iqz}\, (2\pi)^4 \delta^{(4)} (k-p-q)
		\mathcal{M}^{cab}(p,q) 
\label{qcc1}
\end{align}
where
the Wick contractions for evaluating the correlator $\langle J^c(x) J^a (y) J^b(z) \rangle$
in momentum space\footnote{
We Fourier transform the three-point function $\langle J^c(x) J^a (y) J^b(z) \rangle$  by
 \begin{equation}
  \int d^4 x   \int d^4 y \int d^4 z\,   e^{ikx} e^{-ipy} e ^{-iqz} \langle J^c(x) J^a (y) J^b(z) \rangle 
 = (2\pi)^4 \delta^{(4)} (k-p-q)\mathcal{M}^{cab}(p,q)
 \nonumber
\end{equation} 
 with incoming momentum $k$ at vertex $J^c$, and outgoing momenta $p$ and $q$ 
 at vertices $J^a$ and $J^b$, respectively.} 
give rise to the  usual triangle diagrams
\begin{equation}
\begin{tikzpicture} [baseline=(current bounding box.center)] 
\begin{feynman}
\vertex (a);
\vertex [above right= of a] (b);
\vertex [below right= of a] (c);
\vertex [left = of a] (i1) {\( c\)};
\vertex [right= of b] (f1) {\( a\)};
\vertex [right= of c] (f2) {\(b\)};

\diagram* {
	(i1) -- [boson, momentum=\(k\)] 
	(a) -- [anti fermion, edge label= \(l -p\)]  (b) -- [boson, momentum=\(p\)] (f1) [ particle= \(e\) ],
	
	(a) -- [fermion, edge label'=\(l + q\)] (c) --[fermion, edge label=\(l\)] (b),
	(c) -- [boson, momentum'=\(q\)] (f2),
};  
\end{feynman}
\end{tikzpicture}
\quad + \quad
\begin{tikzpicture} [baseline=(current bounding box.center)]
\begin{feynman}
\vertex (a);
\vertex [above right= of a] (b);
\vertex [below right= of a] (c);
\vertex [left = of a] (i1) {\( c\)};
\vertex [right= of b] (f1) {\( b\)};
\vertex [right= of c] (f2) {\(a\)};

\diagram* {
	(i1) -- [boson, momentum=\(k\)] (a) -- [ anti fermion, edge label= \(l - q\)]  
	(b) -- [boson, momentum=\(q\)] (f1) [ particle= \(e\) ],
	
	(a) -- [fermion, edge label'=\(l + p\)] (c) --[ fermion, edge label=\(l\)] (b),
	(c) -- [boson, momentum'=\(p\)] (f2),
};  
\end{feynman}
\end{tikzpicture}
\end{equation}
and produce  the  expression
\begin{equation} 
\begin{split}
\mathcal{M}^{c a b} (p, q)= \ & - i \int \frac{d^n l}{(2 \pi)^n} \frac {{\rm tr} \left\lbrace \gamma^c P_L (\slashed{l} -\slashed{p}) P_R \gamma^a P_L \slashed{l} P_R \gamma^b P_L (\slashed{l} +\slashed{q})  P_R   \right\rbrace } {l^2 (l -p)^2  (l+q)^2} 
+ \Big ( (p,a) \leftrightarrow (q,b) \Big ) \;.
\end{split}
\end{equation}
External momenta are kept four-dimensional, while the
loop momentum $l$ has been extended to $n$ dimensions by dimensional regularization.
It splits into a four-dimensional part (denoted by a bar) and a $(n-4)$-dimensional part (denoted by a hat), as discussed in appendix \ref{B},
i.e. $l= \bar{l} + \hat{l}= \bar{l} + s$ 
(we find it notationally easier to denote the $(n-4)$-dimensional part by $\hat{l}= s$).
Taking the divergence of eq. \eqref{qcc1} is equivalent to contract the above expression 
by $-i k_c$. Considering cyclicity of the trace, we write it as
\begin{equation}
\begin{split}
- i k_c \mathcal{M}^{c a b} (p, q)=\ & - \int \frac{d^n l}{(2 \pi)^n} \frac {{\rm tr}\left\lbrace P_R \slashed{k} 
P_L (\slashed{l} -\slashed{p}) P_R \gamma^a P_L \slashed{l} P_R \gamma^b 
P_L (\slashed{l} +\slashed{q}) \right\rbrace } {l^2 (l -p)^2  (l+q)^2}
+ \Big ( (p,a) \leftrightarrow (q,b) \Big ) 
\end{split}
\end{equation}
and using the property $P_R \gamma^a P_L = \bar{ \gamma}^a P_L$, 
which further enforces the indices $a$, $b$, $c$ to be four-dimensional
(as they contract with external four-dimensional momenta), leads to 
\begin{equation}
\begin{split}
- i k_c \mathcal{M}^{c a b} (p, q)=\ & - \int \frac{d^n l}{(2 \pi)^n} \frac {{\rm tr} \left\lbrace  \slashed{k} 
P_L (\slashed{l} -\slashed{p})  \bar{\gamma}^a P_L \slashed{l} \bar{\gamma}^b 
P_L (\slashed{l} +\slashed{q}) \right\rbrace } {l^2 (l -p)^2  (l+q)^2} + \Big ( (p,a) \leftrightarrow (q,b) \Big ) \;.
\end{split}
\end{equation}
The above expression contains sixteen terms, obtained by expanding the projectors,
but most of them cancel pairwise. Indeed, 
considering first parity-even terms, we may use\footnote{
As explained in appendix \ref{B}, in parity-even calculations one can use a fully anticommuting $\gamma^5$  without running into any mathematical inconsistency.} 
 $\left\lbrace \gamma^5, \gamma^a\right\rbrace =0$ 
together with $(\gamma^5)^2= \mathbb{1}$, so that all parity-even terms become equal to
\begin{equation}
 \int \frac{d^n l}{(2 \pi)^n} \frac{ {\rm tr} \left\lbrace \slashed{k}(\slashed{l} - 
 \slashed{p}) \bar{\gamma}^a \slashed{l} \bar{\gamma}^b(\slashed{l} + \slashed{q})\right\rbrace }{l^2 (l-p)^2 (l+q)^2}  
  + \int \frac{d^n l}{(2 \pi)^n} \frac{ {\rm tr} \left\lbrace \slashed{k} (\slashed{l} 
  - \slashed{q}) \bar{\gamma}^b \slashed{l} \bar{\gamma}^a (\slashed{l} + \slashed{p})\right\rbrace }{l^2 (l-q)^2 (l+p)^2} \ .
\end{equation}
Then, using the identity $\slashed{k} = \slashed{p} + \slashed{q} = \slashed{l} + \slashed{q} - (\slashed{l} - \slashed{p})$ 
in the first integral, and $\slashed{p} + \slashed{q} = \slashed{l} + \slashed{p} - (\slashed{l} - \slashed{q})$ 
in the second one, allows to remove one of the propagators from the integrands. 
The remaining terms cancel out pairwise after changing integration variable $l \to - l$ and using cyclicity of the trace.
Parity-odd terms can be simplified in a similar way, using the following identities
\begin{align}
& \slashed{k} \gamma^5=-\gamma^5 \slashed{k}=- \gamma^5(\slashed{p} + \slashed{q})
= \gamma^5(\slashed{l} - \slashed{p}) + (\slashed{l} + \slashed{q})\gamma^5 - 2 \gamma^5 \slashed{s} \\[1mm]
& \slashed{k} \gamma^5 =-\gamma^5 \slashed{k}=- \gamma^5(\slashed{p} + \slashed{q})
= \gamma^5(\slashed{l} - \slashed{q}) + (\slashed{l} + \slashed{p})\gamma^5 - 2 \gamma^5 \slashed{s} \ .
\end{align} 
At the end one is left with the terms
\begin{align}
- i k_c \mathcal{M}^{c a b} (p, q)=\ & \frac{1}{8} \int \frac{d^n l}{(2 \pi)^n} \frac { {\rm tr} 
\big \lbrace  \gamma^5 \slashed{s} (\bar{\slashed{l}}  + \slashed{s} -\slashed{p})  \bar{\gamma}^a (\bar{\slashed{l}} 
+ \slashed{s}) \bar{\gamma}^b  (\bar{\slashed{l}} + \slashed{s} +\slashed{q})
 \big\rbrace } {l^2 (l -p)^2  (l+q)^2}
\nonumber
 \\
& + \frac{1}{8} \int \frac{d^n l}{(2 \pi)^n} \frac { {\rm tr} 
\big\lbrace  \gamma^5 \slashed{s} (\bar{\slashed{l}}  + \slashed{s} -\slashed{p}) 
 \bar{\gamma}^a (\bar{\slashed{l}} - \slashed{s}) \bar{\gamma}^b  (\bar{\slashed{l}} 
 + \slashed{s} +\slashed{q}) \big\rbrace } {l^2 (l -p)^2  (l+q)^2} 
\nonumber
\\
& + \Big ( (p,a) \leftrightarrow (q,b) \Big ) 
\end{align}
where in the second line the properties of $\gamma^5$ have been used to bring two $\gamma^5$ matrices together,
leading effectively to the replacement of the factor 
$\bar{l} + \slashed{s}$ by $\bar{l} - \slashed{s}$, thus relating the second line to the first one. 
Let us focus on these first two integrals, and rewrite the denominator in symmetric form by using the Feynman parametric formula
\begin{equation}
	\frac{1}{l^2 (l -p)^2  (l+q)^2} = 2 \int_0^1 dx \int_0 ^{1-x} dy \ \Big( \bar{r}^2 + s^2 + f \Big)^{-3}
\end{equation}
where  $\bar{r}=\bar{l} + xq - yp$ is four dimensional, while 
$f\equiv f(x,y,p,q)=   xq^2 + yp^2 - (xq-yp)^2$. 
Since the denominator is symmetric in the internal momentum $s$, 
only terms having an even number of $s$ 
can contribute, otherwise the integral vanishes by symmetric integration. 
Moreover, a term like $ {\rm tr} \left( \gamma^5 \slashed{s} \slashed{s} \bar{\gamma}^a\slashed{s} \bar{\gamma}^b \slashed{s}\right) = s^4 {\rm tr} \left( \gamma^5 \bar{\gamma}^a \bar{\gamma}^b \right) =0$, 
so that only terms proportional to $s^2$ may give a nonvanishing result. 
These terms contain traces of $\gamma^5$ with four-dimensional gamma matrices 
and can be evaluated directly.
After shifting\footnote{We assume that one may shift the momentum integration variable in dimensional regularization, 
and take it as a defining property of dimensional regularization, as discussed for example in \cite{Collins:1984xc}.}
the integration variable $\bar l \to \bar r $, neglecting terms linear in $\bar{r}$ 
that vanish by symmetric integration
and terms that vanish by contraction with the antisymmetric Levi-Civita symbol, one obtains for the numerator
\begin{equation}
	- 8 i s^2 \epsilon^{a b c d}p_c q_d (x + y) \ .
\end{equation}
Integration over $r$ (with measure $d^n r = d^4 \bar r \, d^{n-4} s$, see appendix \ref{C}
for further details) gives a finite result  since
\begin{equation}
\int \frac{d^n r}{(2 \pi)^n} \frac{s^2}{\left( r^2 + f\right)^3 } = -\frac{i}{32 \pi^2}
\end{equation}
 and integration over $x$ and $y$ yields $\frac{2}{3}$. Similar steps hold also for the crossed diagram and lead to an identical result. Finally, we obtain the finite result 
\begin{equation}
	- i k_c \mathcal{M}^{cab} (p, q)= -\frac{8}{3} \frac{1}{32 \pi^2} \epsilon^{a b c d}p_c q_d \ .
\end{equation}
It has to be inserted in the divergence of \eqref{qcc1}, that is
\begin{align}
\partial_c \langle J^c(x) \rangle _{_{\!A}}^{^{\! (2)}}	& = \frac{i}{2} \int d^4 y \int d^4 z\, A_a(y) A_b(z)
\int \frac{d^4 k}{(2 \pi)^4}  \int \frac{d^4 p}{(2 \pi)^4} \int \frac{d^4 q}{(2 \pi)^4} 
\nonumber
 \\ 
& \qquad \qquad  \times e^{-ikx} e^{ipy} e ^{iqz}\, (2\pi)^4 \delta^{(4)} (k-p-q)
\ k_c\mathcal{M}^{cab}(p,q) 
\nonumber
\\ 
& =  \frac{4}{96 \pi^2} \epsilon^{abcd}\int d^4 y \int d^4 z\  A_a(y) A_b(z) \int \frac{d^4 k}{(2 \pi)^4}\int \frac{d^4 p}{(2 \pi)^4} \int \frac{d^4 q}{(2 \pi)^4} 
\nonumber
 \\ 
& \qquad \qquad  \times e^{-ikx}e^{ipy} e ^{iqz}\   p_c q_d \ (2 \pi)^4  \delta^{(4)} (k-p-q)  
\nonumber
 \\ 
& = - \frac{4}{96 \pi^2} \epsilon^{a b c d} \int d^4 y \int d^4 z \Big( \partial_c A_a(y)\Big) \Big( \partial_d A_b(z)\Big) \delta^{(4)}(y-x)\delta^{(4)}(z-x)
\nonumber
\\ 
& =  \frac{4}{96 \pi^2} \epsilon^{abcd} \Big( \partial_a A_b(x)\Big) \Big(\partial_c A_d(x) \Big)  
\nonumber
\\ 
& = \frac{1 }{96\pi^2} \epsilon^{a b c d} F_{ab} (x) F_{cd} (x) 
\end{align}
where in the last line the abelian field strength $F_{ab}= \partial_a A_b - \partial_b A_a$ is introduced. 
This is the correct chiral anomaly, that signals the breakdown of gauge symmetry.
 It prevents a consistent quantization of the gauge field, unless 
other chiral fermions are added to cancel it.

\section{The trace anomaly} 

We now come to consider the quantum properties of the stress tensor, starting with the calculation of the trace anomaly 
in dimensional regularization. The stress tensor of the Weyl fermion 
has been defined in equation \eqref{stress tensor}. It is manifestly symmetric and this property is not modified by 
dimensional regularization.

To derive the trace anomaly, we express in perturbation theory the expectation value 
 $\langle T^{cd}(x)\rangle_{_{\!A}} = \langle T^{cd} (x) e^{iS_{int}} \rangle$ of the stress tensor \eqref{stress tensor}
with $S_{int}=  \int d^4x\, A_a(x) J^a(x) $.
We find it convenient to split the stress tensor in powers of the background field $A_a$ as
  \be
 	T^{cd}  = T_0^{cd}  + T_1^{cd} 
 \ee
 where
\be
 	T_{0}^{cd}  =\frac{1}{4} \bar{\lambda} \big ( 
 	\gamma^c \overset{\leftrightarrow}{\partial^d} + \gamma^d \overset{\leftrightarrow}{\partial^c} \big ) \lambda \;, \qquad
 	T_{1}^{cd}  = -\frac{1}{2} \left( J^c A^d+ J^d A^c\right) 
 \ee
 with the current $J^a$ already defined in \eqref{chiral current}. 
 Then,  at second order in the abelian field $A_a$ we find
\begin{align} 
\langle T^{cd}(x) \rangle_{_{\!A}}^{^{\! (2)}} 
& = -\frac{1}{2} \langle T_{0}^{cd}(x) S_{int} S_{int}\rangle + i \langle T_{1}^{cd}(x) S_{int}\rangle
\nonumber
 \\
&= - \frac{1}{2} \int d^4 y \int d^4 z \, A_a(y) A_b(z) \, \Gamma^{cdab}(x,y,z)
\nonumber 
\\
&=  - \frac{1}{2} \int d^4 y \int d^4 z\, A_a(y) A_b(z)
\int \frac{d^4 k}{(2 \pi)^4}  \int \frac{d^4 p}{(2 \pi)^4} \int \frac{d^4 q}{(2 \pi)^4}  \nonumber
\\
& \quad \quad  \times e^{-ikx} e^{ipy} e ^{iqz}\, (2\pi)^4 \delta^{(4)}(k-p-q)
\mathcal{T}^{cdab}(p,q) 
\label{qst1}
\end{align}
where in the second line
\begin{align}
	\Gamma^{cdab}(x,y,z) &   =   \langle T^{cd}_{0}(x) J^a (y) J^b(z) \rangle 
	\nonumber
	\\
	& \ - \frac{i}{2} \delta^{(4)}(z - x) \left( \eta^{bd}\langle J^c(x) J^a(y) \rangle  + \eta^{bc}\langle J^d(x) J^a(y) \rangle \right)
	\nonumber
	\\
	& \ - \frac{i}{2} \delta^{(4)}(y - x) \left( \eta^{ad}\langle J^c(x) J^b(z) \rangle + \eta^{ac}\langle J^d(x) J^b(z) \rangle \right) \;, 
\end{align}
while in the last line we have Fourier transformed $\Gamma^{cdab}(x,y,z)$ into $\mathcal{T}^{cdab}(p,q)$.
The matrix element $\mathcal{T}^{cdab}(p,q) = \mathcal{T}^{cdab}_{(1)}(p,q) + \mathcal{T}^{cdab}_{(2)}(p,q)$ 
is then given by the following two sets of diagrams
\begin{equation}
\begin{tikzpicture} [baseline=(current bounding box.center)] 
\begin{feynman}
\vertex (a);
\vertex [above right= of a] (b);
\vertex [below right= of a] (c);
\vertex [left = of a] (i1) {\( c d\)};
\vertex [right= of b] (f1) {\( a\)};
\vertex [right= of c] (f2) {\(b\)};

\diagram* {
	(i1) -- [gluon, momentum=\(k\)] (a) -- [ fermion, edge label= \(l \)]  (b) -- [boson, momentum=\(p\)] (f1) [ particle= \(e\) ],
	
	(a) -- [ anti fermion, edge label'=\(l -p -q\)] (c) --[ anti fermion, edge label'=\(l - p\)] (b),
	(c) -- [boson, momentum'=\(q\)] (f2),
};  
\end{feynman}
\end{tikzpicture}
\quad + \quad
\begin{tikzpicture} [baseline=(current bounding box.center)]
\begin{feynman}
\vertex (a);
\vertex [above right= of a] (b);
\vertex [below right= of a] (c);
\vertex [left = of a] (i1) {\( c d\)};
\vertex [right= of b] (f1) {\( b\)};
\vertex [right= of c] (f2) {\(a\)};

\diagram* {
	(i1) -- [gluon, momentum=\(k\)] (a) -- [fermion, edge label= \(l \)]  (b) -- [boson, momentum=\(q\)] (f1) [ particle= \(e\) ],
	
	(a) -- [ anti fermion, edge label'=\(l -p -q\)] (c) --[ anti fermion, edge label'=\(l- q\)] (b),
	(c) -- [boson, momentum'=\(p\)] (f2),
};  
\end{feynman}
\end{tikzpicture}
\end{equation}
\begin{equation}
\begin{tikzpicture} [baseline=(current bounding box.center)] 
\begin{feynman} 

\vertex(a);
\vertex [below=of a] (b);
\vertex [left = of a ] (i1) { \( cd\)};
\vertex [above right  = of a] (f1) { \( b\)};
\vertex [below right= of b] (f2) {\( a\)};

\diagram* {
	
	(i1) -- [gluon, momentum = \(k\)] (a);
	(a) -- [ boson, momentum = \(q\)] (f1);
	(a) -- [  fermion, half left, edge label=\(l \)] (b);
	(b) -- [  fermion,  half left, edge label=\(l-p\)] (a);
	(b)-- [boson, momentum' = \( p\)] (f2);
	
};

\end{feynman}
\end{tikzpicture}
\quad + \quad
\begin{tikzpicture} [baseline=(current bounding box.center)] 
\begin{feynman} 

\vertex(a);
\vertex [below=of a] (b);

\vertex [left = of a ] (i1) { \( cd\)};
\vertex [above right  = of a] (f1) { \( a\)};
\vertex [below right= of b] (f2) {\( b\)};

\diagram* {
	
	(i1) -- [gluon, momentum = \(k\)] (a);
	(a) -- [ boson, momentum = \(p\)] (f1);
	(a) -- [  fermion, half left, edge label=\(l  \)] (b);
	(b) -- [ fermion,  half left, edge label=\(l - q \)] (a);
	(b)-- [boson, momentum' = \( q\)] (f2);
	};
\end{feynman}
\end{tikzpicture}
\end{equation}
leading respectively to the expressions
\begin{equation} \label{integral first diagram}
\mathcal{T}^{cdab}_{(1)} =  - \frac{i}{4} \int \frac{d^n l}{(2 \pi)^n} \frac{\mathcal{N}^{cdab}_{(1)}}{l^2 (l-p)^2 (l-p-q)^2} + \Big( (p,a) \leftrightarrow (q,b) \Big)
\end{equation}
with numerator 
\begin{equation} \label{first numerator}
\begin{split}
\mathcal{N}^{cdab}_{(1)} = {\rm tr} & \left\lbrace  P_R \left( (2l - p -q)^c \gamma^d +(2l - p -q)^d \gamma^c\right) P_L \slashed{l}  P_R\gamma^a P_L (\slashed{l}- \slashed{p}) P_R \gamma^b P_L (\slashed{l} - \slashed{p} - \slashed{q}) \right\rbrace 
\end{split}
\end{equation}	
and 
\begin{equation} \label{integral second diagram}
\mathcal{T}^{cdab}_{(2)} =  \frac{i}{2} \int \frac{d^n l}{(2 \pi)^n}  \frac{ \mathcal{N}^{cdab}_{(2)}  }{l^2(l-p)^2 } + \Big( (p,a) \leftrightarrow (q,b) \Big)
\end{equation}
with
\begin{equation}
\mathcal{N}^{cdab}_{(2)} = {\rm tr} \left\lbrace  P_R\left( \gamma^c \eta^{bd} + \gamma^d \eta^{bc}  \right) P_L \slashed{l} P_R \gamma^a P_L(\slashed{l} - \slashed{p})\right\rbrace .
\end{equation}
The loop momentum $l$ is extended by dimensional regularization as before and the integration measure is $d^n l = d^4 \bar{l} d^{n-4} s$.

Again, using the identities 
$P_R \gamma^a P_L = \bar{ \gamma}^a P_L = P_R \bar{ \gamma}^a $ and $P_L \gamma^a P_R = \bar{ \gamma}^a P_R = P_L \bar{ \gamma}^a$, 
together with the idempotence of the projectors, 
these expressions can
be simplified and the number of projectors reduced to one. Then, the two numerators can be rewritten as
\begin{align}
\mathcal{N}^{cdab}_{(1)} & =  {\rm tr} \left\lbrace  P_R \left(\ (2\bar{l} - p - q)^c \bar{\gamma}^d +(2\bar{l} - p - q)^d \bar{\gamma}^c \right)  \bar{\slashed{l}}  \bar{\gamma}^a (\bar{\slashed{l}} - \slashed{p})  \bar{\gamma}^b (\bar{\slashed{l}} - \slashed{p} - \slashed{q}) \right\rbrace  
\nonumber
\\
& = \frac{1}{2} {\rm tr}  \left\lbrace  \left(  (2\bar{l} - p -q)^c \bar{\gamma}^d +(2\bar{l} - p -q)^d \bar{\gamma}^c \right)  \bar{\slashed{l}} \bar{\gamma}^a (\bar{\slashed{l}}- \slashed{p})  \bar{\gamma}^b (\bar{\slashed{l}} - \slashed{p} -  \slashed{q}) \right\rbrace 
\nonumber
 \\
& \quad - \frac{1}{2} {\rm tr} \left\lbrace  \gamma^5 \left( (2\bar{l} - p -q)^c \bar{\gamma}^d +(2\bar{l}- p -q)^d \bar{\gamma}^c \right) \bar{\slashed{l}}  \bar{\gamma}^a (\bar{\slashed{l}}- \slashed{p})  \bar{\gamma}^b (\bar{\slashed{l}} - \slashed{p} - \slashed{q}) \right\rbrace 
\label{num_1}
\end{align}
and
\begin{align}
\mathcal{N}^{cdab}_{(2)} & =  {\rm tr} \left\lbrace  P_R\left( \bar{\gamma}^c \eta^{bd} + \bar{\gamma}^d \eta^{bc} \right)  \bar{\slashed{l}}   \bar{\gamma}^a (\bar{\slashed{l}} - \slashed{p})\right\rbrace
\nonumber
  \\
& = \frac{1}{2} {\rm tr} \left\lbrace \left( \bar{\gamma}^c \eta^{bd} + \bar{\gamma}^d \eta^{bc}\right)  \bar{\slashed{l}}  \bar{\gamma}^a (\bar{\slashed{l}} - \slashed{p} )\right\rbrace 
\nonumber
 \\
& \quad - \frac{1}{2} {\rm tr} \left\lbrace  \gamma^5 \left( \bar{\gamma}^c \eta^{bd} + \bar{\gamma}^d \eta^{bc}\right)  \bar{\slashed{l}}\bar{\gamma}^a  (\bar{\slashed{l}} - \slashed{p})\right\rbrace .
\label{num_2}
\end{align}
In the above expressions, there is no need of putting a bar over the momenta $p$ and $q$ because, being 
external, they are kept four-dimensional and only the internal momentum $l$ is extended to $n$ dimensions.
We also notice that all gamma matrices are effectively reduced to the four-dimensional ones.

Now, one can show that terms containing traces of $\gamma^5$, which would produce parity-odd contributions, cancel with those of the crossed diagrams. This can be done by using the invariance of the 
trace under transposition, the known properties of the gamma matrices under transposition, namely 
$ C \bar{ \gamma}^a C^{-1} = - \bar{\gamma}^{a\, T} $ and $C \gamma^5 C^{-1} =  \gamma^{5\, T} $
with $C$ the charge conjugation matrix,
a  shift on the integration variable, and the property of the  anticommutator $\left\lbrace \gamma^5, \bar{\gamma}^a\right\rbrace =0$. It proves that parity-odd terms do not appear in the expectation value of the stress tensor, 
and thus in the trace anomaly.
This is a crucial point of our investigation, so let us show in more detail the precise steps we have been
following.

To start with, it is easy to verify that the parity-odd term from eq. \eqref{num_2} vanishes
\begin{equation}
	\int \frac{d^n l}{(2 \pi)^n} \frac{{\rm tr} \left\lbrace  \gamma^5 \left( \bar{\gamma}^c \eta^{bd} + \bar{\gamma}^d \eta^{bc}\right)  \bar{\slashed{l}}\bar{\gamma}^a  (\bar{\slashed{l}} - \slashed{p})\right\rbrace}{l^2 (l - p)^2} = 0
\end{equation}
and similarly for the corresponding crossed term. This implies the vanishing of the parity-odd part 
of \eqref{integral second diagram}.
Now, let us turn to the last line of eq. \eqref{num_1}
that also contains $\gamma^5$ 
\begin{equation}
	\int \frac{d^n l}{(2 \pi)^n} \frac{ {\rm tr} \left\lbrace  \gamma^5 \left( (2\bar{l} - p -q)^c \bar{\gamma}^d +(2\bar{l}- p -q)^d \bar{\gamma}^c \right) \bar{\slashed{l}}  \bar{\gamma}^a (\bar{\slashed{l}}- \slashed{p})  \bar{\gamma}^b (\bar{\slashed{l}} - \slashed{p} - \slashed{q}) \right\rbrace }{l^2 (l-p)^2(l -p-q)^2} + \Big( (p,a) \leftrightarrow (q,b) \Big) \
\;.	 \label{int_2}
\end{equation} 
These terms cancel among themselves, as we shall see immediately. As already noticed, 
the gamma matrices under the trace are now four-dimensional. Then, 
using the invariance of the trace under transposition and the relations
involving the charge conjugation matrix $C$
	\begin{equation}
	C \bar{ \gamma}^a C^{-1} = - \bar{\gamma}^{a\, T} \ ,  \qquad C \gamma^5 C^{-1} =  \gamma^{5\, T} \;,  
	\end{equation}
one can write
\begin{align}
& (2\bar{l} - p -q)^c \ {\rm tr} \left\lbrace  \gamma^5 \bar{\gamma}^d  \bar{\slashed{l}}  \bar{\gamma}^a (\bar{\slashed{l}}- \slashed{p})  \bar{\gamma}^b (\bar{\slashed{l}} - \slashed{p} - \slashed{q}) \right\rbrace 
\nonumber
\\
& = (2\bar{l} - p -q)^c \ {\rm tr} \left\lbrace  \gamma^5 \bar{\gamma}^d  \bar{\slashed{l}}  \bar{\gamma}^a (\bar{\slashed{l}}- \slashed{p})  \bar{\gamma}^b (\bar{\slashed{l}} - \slashed{p} - \slashed{q}) \right\rbrace ^ T 
\nonumber
\\
& =  (2\bar{l} - p -q)^c \ {\rm tr} \left\lbrace (\bar{\slashed{l}} - \slashed{p} - \slashed{q})^T ( \bar{\gamma}^b)^T (\bar{\slashed{l}}- \slashed{p})^T (\bar{\gamma}^a)^T (\bar{\slashed{l}})^T (\bar{\gamma}^d )^T \left( \gamma^5\right)^T   \right\rbrace 
\nonumber
\\
& =  (2\bar{l} - p -q)^c \ {\rm tr} \left\lbrace (\bar{\slashed{l}} - \slashed{p} - \slashed{q})  \bar{\gamma}^b (\bar{\slashed{l}}- \slashed{p}) \bar{\gamma}^a \bar{\slashed{l}} \bar{\gamma}^d  \gamma^5   \right\rbrace 
\nonumber
\\
& = -  (2\bar{l} - p -q)^c \ {\rm tr} \left\lbrace \gamma^5 \bar{\gamma}^d (\bar{\slashed{l}} - \slashed{p} - \slashed{q})  \bar{\gamma}^b (\bar{\slashed{l}}- \slashed{p}) \bar{\gamma}^a \bar{\slashed{l}} \right\rbrace
\end{align}
where in the last line the anticommutator $\left\lbrace \gamma^5, \bar{ \gamma}^a\right\rbrace = 0 $ has been used. Following similar steps one finds 
\begin{equation}
	(2\bar{l} - p -q)^d \ {\rm tr} \left\lbrace  \gamma^5 \bar{\gamma}^c  \bar{\slashed{l}}  \bar{\gamma}^a (\bar{\slashed{l}}- \slashed{p})  \bar{\gamma}^b (\bar{\slashed{l}} - \slashed{p} - \slashed{q}) \right\rbrace =  -  (2\bar{l} - p -q)^d \ {\rm tr} \left\lbrace \gamma^5 \bar{\gamma}^c (\bar{\slashed{l}} - \slashed{p} - \slashed{q})  \bar{\gamma}^b (\bar{\slashed{l}}- \slashed{p}) \bar{\gamma}^a \bar{\slashed{l}} \right\rbrace 
\end{equation}
and \eqref{int_2} becomes
\begin{align}
 & - \int \frac{d^n l}{(2 \pi)^n} \frac{{\rm tr} \left\lbrace \gamma^5 \left( (2\bar{l} - p -q)^c \bar{\gamma}^d + (2\bar{l} - p -q)^d\bar{\gamma}^c\right)  (\bar{\slashed{l}} - \slashed{p} - \slashed{q})  \bar{\gamma}^b (\bar{\slashed{l}}- \slashed{p}) \bar{\gamma}^a \bar{\slashed{l}} \right\rbrace }{l^2 (l-p)^2(l-p-q)^2} 
 \nonumber
 \\
 & + \int \frac{d^n l}{(2 \pi)^n} \frac{{\rm tr} \left\lbrace  \gamma^5 \left( (2\bar{l} - p -q)^c \bar{\gamma}^d +(2\bar{l}- p -q)^d \bar{\gamma}^c \right) \bar{\slashed{l}}  \bar{\gamma}^b (\bar{\slashed{l}} - \slashed{q})  \bar{\gamma}^a (\bar{\slashed{l}} - \slashed{p} - \slashed{q}) \right\rbrace}{l^2 (l-q)^2(l-p-q)^2}
 \end{align}
where in the second line the crossed term has been made explicit. After shifting the integration variable of the first integral as\footnote{The change of variable can also be written as $\bar{l} \to - \bar{l} + p +q,\  s \to -s$.}  $l \to -l + p + q$, one finally obtains
\begin{align}
& - \int \frac{d^n l}{(2 \pi)^n} \frac{{\rm tr} \left\lbrace  \gamma^5 \left( (2\bar{l} - p -q)^c \bar{\gamma}^d +(2\bar{l}- p -q)^d \bar{\gamma}^c \right) \bar{\slashed{l}}  \bar{\gamma}^b (\bar{\slashed{l}} - \slashed{q})  \bar{\gamma}^a (\bar{\slashed{l}} - \slashed{p} - \slashed{q}) \right\rbrace}{l^2 (l-q)^2(l-p-q)^2} 
\nonumber
\\
& + \int \frac{d^n l}{(2 \pi)^n} \frac{{\rm tr} \left\lbrace  \gamma^5 \left( (2\bar{l} - p -q)^c \bar{\gamma}^d +(2\bar{l}- p -q)^d \bar{\gamma}^c \right) \bar{\slashed{l}}  \bar{\gamma}^b (\bar{\slashed{l}} - \slashed{q})  \bar{\gamma}^a (\bar{\slashed{l}} - \slashed{p} - \slashed{q}) \right\rbrace}{l^2 (l-q)^2(l-p-q)^2} = 0 \ .
\end{align}
This proves the vanishing of the parity-odd part of \eqref{integral first diagram}. Thus, 
parity-odd terms do not appear in the expectation value of the stress tensor and in the 
corresponding trace anomaly.
Similar manipulations with the charge conjugation matrix have been used in \cite{Armillis:2010pa}.
One could interpret physically these algebraic results by recalling the CP invariance of the 
unregulated Weyl theory, which describes a spin 1/2  particle composed of two states, interpretable
 as a left handed particle plus its right handed antiparticle, as well-known for the original Weyl theory of the massless neutrino.
 These two states of the  Weyl theory have chiral effects that balance each other out in the trace of the stress tensor.

Now, as explained in appendix \ref{B}, in parity-even calculations one can use a fully anticommuting $\gamma^5$  without running into any mathematical inconsistency. 
This implies that in the above expressions of the numerators $\mathcal{N}^{cdab}_{(1)}$ and $\mathcal{N}^{cdab}_{(2)}$ the four-dimensional gamma matrices can be replaced by $n$-dimensional ones, that is
\begin{equation} 
\mathcal{N}^{cdab}_{(1)} = \frac{1}{2} {\rm tr}  \left\lbrace \left(  (2l - p - q)^c \gamma^d +(2l - p - q)^d \gamma^c \right)  \slashed{l}  \gamma^a (\slashed{l} - \slashed{p})  \gamma^b  (\slashed{l} - \slashed{p} - \slashed{q}) \right\rbrace 
\end{equation}
and
\begin{equation}
\mathcal{N}^{cdab}_{(2)} = \frac{1}{2} {\rm tr} \left\lbrace  \left( \gamma^c \eta^{bd} + \gamma^d \eta^{bc} \right) \slashed{l} \gamma^a (\slashed{l} - \slashed{p})\right\rbrace \ .
\end{equation}

To obtain the trace anomaly, we contract the expectation value \eqref{qst1} with the four-dimensional metric tensor $\bar{\eta}^{ab}$
\begin{align} \label{qt1}
\langle T^a_{\ a} (x) \rangle _{_{\!A}}^{^{\! (2)}} \equiv \bar{\eta}_{cd}\langle T^{cd}(x) \rangle_{_{\!A}}^{^{\! (2)}} 
& =  - \frac{1}{2} \int d^4 y \int d^4 z\, A_a(y) A_b(z)
\int \frac{d^4 k}{(2 \pi)^4}  \int \frac{d^4 p}{(2 \pi)^4} \int \frac{d^4 q}{(2 \pi)^4}
\nonumber 
\\
& \quad \quad  \times e^{-ikx} e^{ipy} e ^{iqz}\, (2\pi)^4 \delta^{(4)}(k-p-q)
\mathcal{T}^{ab}(p,q) 
\end{align}
where
\begin{align} \label{qt2}
\mathcal{T}^{ab} (p, q) \equiv \bar{\eta}_{cd} \mathcal{T}^{cdab} (p, q) = & \ - \frac{i}{4} 
\int \frac{d^n l}{(2 \pi)^n} 
\frac{{\rm tr} \left\lbrace (2 \slashed{\bar{l}} - \slashed{p} - \slashed{q}) \slashed{l} \gamma^a (\slashed{l} - \slashed{p}) \gamma^b (\slashed{l} - \slashed{p} - \slashed{q}) \right\rbrace}
{l^2(l-p)^2 (l-p-q)^2} 
\nonumber
\\
& \quad + \frac{i}{2} \int \frac{d^n l}{(2 \pi)^n} \frac{{\rm tr}\left\lbrace  \gamma^b \slashed{l} \gamma^a (\slashed{l} - \slashed{p}) \right\rbrace }{l^2 (l-p)^2} 
\nonumber
\\
& \quad + \Big( (p,a) \leftrightarrow (q,b)\Big) \ .
\end{align}

Of course,  $\gamma^a$ and $\gamma^b$ remain essentially four-dimensional as they are contracted 
with external gauge fields which are kept four-dimensional. This expression is further simplified by rewriting $\slashed{\bar{l}} = \slashed{l} - \slashed{s}$ and using the identity
\begin{equation} \label{id_1}
\frac{ (\slashed{l} - \slashed{p} - \slashed{q}) (2 \slashed{l} - \slashed{p} - \slashed{q}) }{l^2 (l-p)^2(l- p- q)^2}=\frac{2}{l^2 (l-p)^2} + \frac{ (\slashed{l} - \slashed{p} - \slashed{q})(\slashed{p} + \slashed{q}) }{l^2 (l-p)^2 (l-p -q)^2} \ .
\end{equation}
From the first term of the right-hand side we get an expression which cancels the integral in the second line of \eqref{qt2}. A similar identity can be written for the crossed terms. Moreover, the second term of the above identity gives a term that cancels with that one of the crossed diagram. At the end, one is left with the expression
\begin{equation}
\mathcal{T}^{ab} (p,q)= \frac{i}{2} \int \frac{d^n l}{(2 \pi)^n} \frac{ {\rm tr} \left\lbrace \slashed{s}\slashed{l} \gamma^a (\slashed{l} - \slashed{p})  \gamma^b  (\slashed{l} - \slashed{p} - \slashed{q})\right\rbrace }{l^2 (l-p)^2 (l-p -q)^2} + \Big( (p,a) \leftrightarrow (q,b) \Big) \ .
\end{equation}
Using the Feynman parametric formula to rewrite the denominator in symmetric form, 
see appendix \ref{C}, the integral becomes
\begin{equation}
i \int_{0}^{1} dx \int_{0}^{1-x} dy \int \frac{d^4 \bar{l}}{(2 \pi)^4} \int \frac{d^{n-4} s}{(2 \pi)^{n-4}} \ \frac{N^{ab}}{\left( \bar{l}^2 + s^2 + f\right)^3 } 
\end{equation}
where the integration variable has been shifted by $\bar{l} \to \bar{l} + p(x+y) + qx$, 
$f\equiv f(x,y,p,q)=p^2[x(1-x) + y(1-y) - 2xy] + x(1-x)q^2 + 2 p qx (1-x-y)$, and the numerator becomes
\begin{equation}
N^{ab}  = {\rm tr} \left\lbrace \slashed{s} (\slashed{\bar{l}} + \slashed{s} + \slashed{p}(x+y) + \slashed{q}x) \gamma^a (\slashed{\bar{l}} + \slashed{s} + \slashed{p}(x+y-1) + \slashed{q}x)  \gamma^b (\slashed{\bar{l}} + \slashed{s} + \slashed{p}(x+y-1) + \slashed{q}(x-1)) \right\rbrace   \ .
\end{equation}
By symmetry only terms proportional to even powers of $s$ give a nonvanishing contribution. 
There are three terms proportional to $s^2$ and one proportional to $s^4$. After working the traces out, neglecting linear term in $\bar{l}$, replacing $\bar{l}^a \bar{l}^b = \frac{1}{4} \bar{\eta}^{ab}{\bar{l}}^2$, and evaluating the loop integrals using 
 \begin{align} 
 & \int \frac{d^4 \bar{l}}{(2 \pi)^4} \int \frac{d^{n-4} s}{(2 \pi)^{n-4}} \frac{s^2}{(\bar{l}^2 + s^2 + f)^3}= - \frac{i}{32 \pi^2}  \\
 & \int \frac{d^4 \bar{l}}{(2 \pi)^4} \int \frac{d^{n-4} s}{(2 \pi)^{n-4}} \frac{s^4}{(\bar{l}^2 + s^2 + f)^3}
 =  \frac{i}{32 \pi^2} f \;, 
 \end{align}
we obtain 
\begin{equation}
\frac{1}{8 \pi^2} \int_0^1dx \int_0^{1-x} dy \ \bar{N}^{ab}
\end{equation}
where 
\begin{equation}
\begin{split}
\bar{N}^{ab} & =  2 p^a p^b (2x^2 + 2y^2 + 4xy - 3x - 3y +1) + p^a q^b (4x^2 + 4xy - 4x -2y +1)  \\
& \quad +  q^a p^b (4x^2 + 4xy - 4x +1) + q^a q^b (4x^2-2x)  \\
& \quad - \eta^{ab} \Big(  p^2 (2x^2 + 4xy -3x + 2y^2 - 3y +1) +  q^2 (2x^2 -x) + p  q (4x^2 + 4xy - 4x +1)\Big)  \ .
\end{split}
\end{equation}
Integrating over $x$ and $y$, the only contribution is from $\int_0^1 dx \int_0^{1-x} dy \ (4x^2 + 4xy - 4x +1)= \frac{1}{3}$. Similar steps hold also for the crossed terms and lead to an identical result, so that at the end we find
\begin{equation}
	\mathcal{T}^{ab} (p,q) = \frac{1}{12 \pi^2} \left( q^a p^b - \eta^{ab} pq\right) \ .  
\end{equation}
This result has to be inserted in \eqref{qt1} and gives the final expression of the trace anomaly
\begin{align}
\langle T^a_{\ a}(x)\rangle _{_{\!A}}^{^{\! (2)}} & = - \frac{1}{2} \int d^4y \int d^4 z\  A_a(y) A_b(z) \int \frac{d^4 k}{(2 \pi)^4} \int \frac{d^4 p}{(2 \pi)^4} \int \frac{d^4 q}{(2 \pi)^4} 
\nonumber
\\
& \qquad \qquad \times e^{-ikx} e^{ipy} e^{iqz} (2 \pi)^4 \delta^{(4)}(k-p-q)\mathcal{T}^{ab}(p,q)
\nonumber
\\
& = - \frac{1}{24 \pi^2} \int d^4y \int d^4 z \ A_a(y) A_b(z)\int \frac{d^4 k}{(2 \pi)^4}  \int \frac{d^4 p}{(2 \pi)^4} \int \frac{d^4 q}{(2 \pi)^4} 
\nonumber
\\
& \qquad \qquad \times e^{ - ikx} e^{ipy} e^{iqz} (2 \pi)^4 \delta^{(4)}(k-p-q) \Big(  q^a p^b - \eta^{ab}p q \Big) 
\nonumber
\\
& = - \frac{1}{24 \pi^2} \int d^4y \int d^4 z \left( \eta^{ab} \partial_c^y \partial^c_z - \partial^a_z \partial^b_y\right)  A_a(y) A_b(z) \times 
\nonumber
\\
& \qquad \qquad  \times\int \frac{d^4 k}{(2 \pi)^4}\int \frac{d^4 p}{(2 \pi)^4} \int \frac{d^4 q}{(2 \pi)^4} e^{-ikx} e^{ipy} e^{iqz} (2 \pi)^4 \delta^{(4)}(k-p-q) 
\nonumber
\\
& = - \frac{1}{24 \pi^2} \int d^4y \int d^4 z \Big[  \Big(  \partial_a A_b(y) \Big)  \Big(  \partial^a A^b(z) \Big)  - \Big(  \partial_a A_b(z) \Big) \Big(  \partial^b A^a(y) \Big) \Big]  \delta ^{(4)}(y-x) \delta^{(4)} (z-x)
\nonumber
\\
& = - \frac{1}{24 \pi^2} \Big[ \Big( \partial_a A_b(x) \Big) \Big( \partial^a A^b(x) \Big) - \Big( \partial_a A_b (x) \Big) \Big( \partial^b A^a(x) \Big)  \Big] 
\nonumber
\\
& = - \frac{1}{48 \pi^2} F^{ab}(x)F_{ab}(x) \ .
\end{align}
This is the trace anomaly of a chiral fermion \cite{Bastianelli:2018osv}.
As one may check, it corresponds precisely to half the trace anomaly of a Dirac fermion.
However, we cannot yet assert with certainty that this is the exact expression of the trace anomaly 
of a Weyl fermion
without first checking the conservation of the stress tensor. 
That is because the Breitenlohner-Maison/'t Hooft-Veltman 
scheme breaks $n$-dimensional Lorentz covariance of chiral theories 
and the conservation of the stress tensor is not guaranteed at the quantum level. 
As a consequence, it may happen that to preserve the conservation one needs to introduce 
counterterms\footnote{Conservation is guaranteed, as there are no
genuine gravitational anomalies in four dimensions \cite{Alvarez-Gaume:1983ihn}.},
which in turn may modify the expression of the trace anomaly.
As we shall see, this is not the case.

\section{Stress tensor conservation}
The Ward identity associated with the conservation of the stress tensor,
corresponding to equation \eqref{c2}, is
\begin{equation}
	\partial_c \langle T^{cd} \rangle_{_{\!A}} = - F^{cd}  \langle J_c \rangle_{_{\!A}} \ .
	\label{45}
\end{equation}
To check that there are no anomalies associated with this equation 
we expand both sides of this equation at second order in the background gauge field, and verify 
their equality\footnote{In \cite{Giannotti:2008cv} and \cite{Armillis:2009pq} this strategy was used to determine the structure of the Ward identity for the conservation of the stress tensor of a Dirac fermion. We do the same thing, but for a Weyl fermion.}.
For the left-hand side we get
\begin{align} 
\partial_c\langle T^{cd}(x) \rangle_{_{\!A}}^{^{\! (2)}}  
 &= - \frac{1}{2} \int d^4 y \int d^4 z \, A_a(y) A_b(z) \, \partial_c^x \  \Gamma^{cdab}(x, y, z)
 \nonumber 
 \\
& =  - \frac{1}{2} \int d^4 y \int d^4 z\, A_a(y) A_b(z)
\int \frac{d^4 k}{(2 \pi)^4}  \int \frac{d^4 p}{(2 \pi)^4} \int \frac{d^4 q}{(2 \pi)^4}  \nonumber
\\
& \quad \quad  \times e^{-ikx} e^{ipy} e ^{iqz}\, (2\pi)^4 \delta^{(4)} (k-p-q)
\left( - i k_c \right) \mathcal{T}^{cdab}(p,q) \ .
\end{align}
By expanding perturbatively the right-hand side of equation \eqref{45} one finds
\begin{equation}
	- F^{cd}\langle J_c \rangle_{_{\!A}} = - F^{cd} \langle J_c \rangle - i F^{cd} \langle J_c S_{int} \rangle + . . .
\end{equation}
where $S_{int} = \int d^4 x A_a(x)J^a(x)$. Only the second term of the above expansion contributes, since the first one vanishes, and at second order in the background gauge field it gives
\begin{align}
	- F^{cd} (x) \langle J_c (x) \rangle_{_{\!A}}^{(2)} & = \frac{i}{2} \int d^4 y \int d^4 z \ A_a(y) A_b(z) \  \times
	\nonumber
	\\
	  & \quad  \times \Big[   \left( \eta^{bd} \delta^e_c - \delta^b_c \eta^{de}\right) \langle J^c(x) J^a(y) \rangle \ \partial_e^z \  \delta^{(4)} (z-x)
	\nonumber
	\\ 
	&  \qquad +  \left( \eta^{ad} \delta^e_c - \delta^a_c \eta^{de}\right) \langle J^c(x) J^b(z) \rangle \  \partial_e^y \  \delta^{(4)} (y - z ) \Big]	\;.
\end{align}
By comparing both sides of equation \eqref{45} we obtain the identity 
\begin{align}
\partial_c^x \Gamma^{cdab} (x,y,z) & = - i \left( \eta^{bd} \delta^e_c - \delta^b_c \eta^{de}\right) \langle J^c(x) J^a(y) \rangle \ \partial_e^z \  \delta^{(4)} (z-x)
\nonumber
\\
& \quad  - i \left( \eta^{ad} \delta^e_c - \delta^a_c \eta^{de}\right) \langle J^c(x) J^b(z) \rangle \  \partial_e^y \  \delta^{(4)} (y - z ) 
\end{align}
which in momentum space becomes
\begin{equation} \label{msc2}
- i k_c \mathcal{T}^{cdab} (p,q) = \left( \eta^{bd} q_c - \delta^b_c q^d \right)  \Pi^{ca} (p) + \left( \eta^{ad} p_c - \delta^a_c p^d \right) \Pi^{cb} (q)
\end{equation}
where $\Pi^{ab}(p)$ is the Fourier transform of the two-point function $\langle J^a(x) J^b(y)\rangle$, 
whose expression is
\begin{equation}
	\Pi^{ab} (p) = \int \frac{d^n l}{(2 \pi)^n} \frac{{\rm tr}\left\lbrace \gamma^a P_L \slashed{l} P_R \gamma^b P_L (\slashed{l} - \slashed{p}) P_R\right\rbrace }{l^2(l-p)^2} 
\end{equation}
where the integration variable $l$ has been extended by dimensional regularization.
It is easy to see that this integral reduces to half the one of the vacuum polarization (photon self-energy), i.e.
\begin{align}
\Pi ^{ab} (p) & = \frac{1}{2} \int \frac{d^n l}{(2 \pi)^n} \frac{{\rm tr}\left\lbrace \gamma^a \slashed{l} \gamma^b (\slashed{l} - \slashed{p})\right\rbrace }{l^2(l-p)^2} 
\nonumber
\\ 
& \overset{n \to 4}{=} - \frac{i}{24 \pi^2} \left( p^a p^b - \eta^{ab}p^2\right)\left( \frac{2}{4-n} + \frac{5}{3} - \log p^2 - \gamma + \log 4\pi \right)  
\end{align}
where the limit $n \to 4$ has been taken and $\gamma$ is the Euler-Mascheroni constant.
Thus, the right-hand side of equation \eqref{msc2} yields
\begin{align} \label{53}
& \frac{i}{24 \pi^2} \Big(  q^d \left( p^a p^b - \eta^{ab}p^2\right)  +  \eta^{bd}\left(  q^a p^2 -  p^a p q \right)  \Big)  \left( \frac{2}{4-n} + \frac{5}{3}- \log p^2 - \gamma + \log 4\pi + O(4-n)\right) +
\nonumber
 \\
& \ +  \frac{i}{24\pi^2} \Big(  p^d \left( q^a q^b - \eta^{ab}q^2\right) + \eta^{ad} \left( p^b q^2 - q^b p q \right)  \Big) \left( \frac{2}{4-n} + \frac{5}{3} - \log q^2 - \gamma + \log 4\pi + O(4-n) \right) \ .
\end{align}
Let us now evaluate the left-hand side of equation \eqref{msc2}
\begin{align} 
- i  k_c \mathcal{T}^{cdab} (p, q) & =  - \frac{1}{8} \int \frac{d^n l}{(2 \pi)^n} \frac{ {\rm tr} \left\lbrace \left[ k_c (2l - k)^c \gamma^d +(2l - k)^d \slashed{k} \right]  \slashed{l}  \gamma^a (\slashed{l} - \slashed{p})  \gamma^b  (\slashed{l} - \slashed{k}) \right\rbrace }{l^2 (l-p)^2 (l-k)^2} 
\nonumber
\\
& \quad + \frac{1}{4} \int \frac{d^n l}{(2 \pi)^n}  \frac{ {\rm tr} \left\lbrace \left( \slashed{k} \eta^{bd} + \gamma^d k^b \right) \slashed{l}  \gamma^a  (\slashed{l} - \slashed{p}) \right\rbrace  }{l^2 (l-p)^2} 
\nonumber
\\
&  \quad + \Big( (p,a) \leftrightarrow (q,b) \Big) 
\end{align}
where $k = p +q$ because of momentum conservation.
The following two identities can be used to simplify the calculation
\begin{align}
& k (2l-k)= 2 l k - k^2 = l^2 -(l-k)^2 \\
& \slashed{k}= \slashed{l} - (\slashed{l} - \slashed{k})
\end{align}
and one has
	\begin{align}
	- i k_c \mathcal{T}^{cdab} (p,q) & = - \frac{1}{4} \int \frac{d^n l}{(2 \pi)^n} {\rm tr} \left\lbrace \gamma^d \slashed{l} \gamma^a (\slashed{l} - \slashed{p}) \gamma^b (\slashed{l} - \slashed{k}) \right\rbrace \left( \frac{1}{(l-p)^2 (l-k)^2} - \frac{1}{l^2 (l-p)^2}\right) 
	\nonumber
	\\
	& \quad	- \frac{1}{4} \int \frac{d^n l}{(2 \pi)^n} \left( 2l -k\right)^d \left(  \frac{{\rm tr} \left\lbrace \gamma^a (\slashed{l} - \slashed{p}) \gamma^b (\slashed{l} - \slashed{k})\right\rbrace }{(l-p)^2 (l-k)^2} - \frac{ {\rm tr} \left\lbrace \slashed{l}\gamma^a (\slashed{l} - \slashed{p})\gamma^b\right\rbrace }{l^2 (l-p)^2}\right) 
	\nonumber
	\\
	& \quad + \frac{1}{4} \ k^b \int \frac{d^n l}{(2 \pi)^n}  \frac{{\rm tr} \left\lbrace \gamma^d \slashed{l}\gamma^a (\slashed{l} - \slashed{p})\right\rbrace }{l^2 (l-p)^2} 
	+ \frac{1}{4} \ \eta^{bd} \int \frac{d^n l}{(2 \pi)^n}  \frac{{\rm tr}\left\lbrace \slashed{q} \slashed{l}\gamma^a (\slashed{l} - \slashed{p})\right\rbrace }{l^2 (l-p)^2} 
	\nonumber
	\\
	& \quad + \frac{1}{4} \ k^a \int \frac{d^n l}{(2 \pi)^n}  \frac{ {\rm tr}\left\lbrace \gamma^d \slashed{l}\gamma^b (\slashed{l} - \slashed{q})\right\rbrace }{l^2 (l-q)^2} 
	 + \frac{1}{4} \ \eta^{ad} \int \frac{d^n l}{(2 \pi)^n}  \frac{{\rm tr} \left\lbrace \slashed{p} \slashed{l}\gamma^b (\slashed{l} - \slashed{q})\right\rbrace }{l^2 (l-q)^2} 
	\end{align}
where the first two integrals have been multiplied by two since those of the crossed diagrams are equal after using the invariance of the trace under transposition and shifting the integration variable. Tadpole integrals have been neglected because they vanish.
The calculation is quite long and we refer to appendix \ref{D} for details. Here we give the final result 
\begin{align} \label{cms}
- i k_c \mathcal{T}^{cdab} (p, q)& = \frac{i}{24 \pi^2} \Bigg( q^d \left( p^a p^b - \eta^{ab}p^2\right)  +  \eta^{bd}\left(  q^a p^2 -  p^a p q \right)  \Bigg) \times 
\nonumber
\\
& \qquad \times \left( \frac{2}{4-n} + \frac{5}{3}- \log p^2 - \gamma + \log 4\pi + O(4-n)\right) 
\nonumber
 \\
& \ +  \frac{i}{24\pi^2} \Bigg(   p^d \left( q^a q^b - \eta^{ab}q^2\right) + \eta^{ad} \left( p^b q^2 - q^b p q \right)  \Bigg)  \times 
\nonumber
\\
&  \qquad \times \left( \frac{2}{4-n} + \frac{5}{3} - \log q^2 - \gamma + \log 4\pi + O(4-n) \right) 
\end{align} 
that matches precisely eq. \eqref{53}, thus verifying
the Ward identity \eqref{45} at quadratic order in the background field $A_a$.

 \section{Conclusions} 

We have used dimensional regularization to study the trace anomaly of a Weyl fermion
coupled to an abelian gauge field, confirming that no parity-odd term arises in its expression, 
as found in \cite{Bastianelli:2018osv} by Pauli-Villars regularization and heat kernel methods.
The resulting expression is gauge invariant, even though the gauge symmetry is anomalous.
It equals half the trace anomaly of a Dirac fermion.
This result matches the analogous case of a chiral fermion in curved spacetime,
which has been much debated in the recent literature, as reviewed in the introduction.
 The coupling to the abelian gauge field--- 
 while interesting in itself---has allowed to expose in a simpler context 
 the subtle points of dimensional regularization of chiral theories, which 
  become much more tedious when the coupling to gravity is turned on.
  We have given a detailed description of the strategy 
  adopted and the steps needed to calculate the trace anomaly.
Our exposition might be useful for comparing with alternative calculations that one may 
wish to adopt in verifying that no parity-odd terms arise in the trace anomaly of a Weyl fermion.

             \vfill\eject

\appendix
\section{Conventions}
\label{A}

We use a mostly plus Minkowski metric $\eta_{ab}$ and gamma matrices satisfying 
\begin{equation}
	\{ \gamma^a, \gamma^b \}= 2 \eta^{ab}\;.
\end{equation}
The hermitian and traceless chiral matrix $\gamma^5$ is defined as
\begin{equation} 
\gamma^5= \frac{i}{4 !} \epsilon_{a b c d} \gamma^a \gamma^b \gamma^c \gamma^d = - i \gamma^0 \gamma^1 \gamma^2 \gamma^3
\label{gamma5}
\end{equation}
with the symbol $\epsilon_{a b c d}$ normalized as $\epsilon_{0 1 2 3}=-1$ and $\epsilon^{0 1 2 3}=1$.
In particular, one has
\begin{equation}
{\rm tr}\,  ( \gamma^5 \gamma^a \gamma^b \gamma^c \gamma^d ) = 4 i \epsilon^{a b c d}.
\end{equation}
The conjugate Dirac spinor $\bar{\lambda}$ is defined using $\beta=i \gamma^0$ by
\begin{equation}
	\bar{\lambda}= \lambda^\dagger \beta \;.
\end{equation}
The charge conjugation matrix $C$ is defined to satisfy 
\begin{equation}
	C \gamma^a C^{-1} = - {\gamma^a}^T \;.
\end{equation}

\section{The Breitenlohner-Maison-'t Hooft-Veltman prescription} 
\label{B}

Dimensional regularization is subtle in chiral theories.
The main problem concerns the extension of purely four-dimensional quantities to $n$ dimensions, such as the chiral matrix $\gamma^5$ 
and the Levi-Civita symbol $\epsilon_{a b c d}$.
In four dimensions we define $\gamma^5$ by eq. \eqref{gamma5}, so that it squares to the identity and 
anticommutes with the other gamma matrices
\begin{equation}
 \{ \gamma^5, \gamma^a \} = 0   \;.
 \label{anticommutator} 
 \end{equation}

In $n$ dimensions there are $n$ gamma matrices satisfying the Clifford algebra
\begin{equation}
	 \{  \gamma^a, \gamma^b  \} = 2 \eta^{ab}, \qquad a,b=0,1,...,n     \quad (a,b\neq 5)    
\end{equation}
and the simplest thing one could try is to extend \eqref{anticommutator} to $n$ dimensions, but a fully 
anticommuting $\gamma^5$ would lead to inconsistency for parity-odd calculations, 
i.e. calculations  involving an odd number of $\gamma^5$ 
matrices \cite{Breitenlohner:1977hr, Baikov:1991qz, Bonneau:1980yb, Gottlieb:1979ix}.

In \cite{tHooft:1972tcz} 't Hooft and Veltman proposed a generalization of $\gamma^5$ 
to $n$ dimensions such that $\gamma^5$ anticommutes with the first four gamma matrices 
and commutes with the remaining $n-4$ matrices, and derived the standard chiral anomaly within this scheme. 
This proposal was further developed by Breitenlohner and Maison \cite{Breitenlohner:1977hr}, 
who proved its consistency to all orders in perturbation theory.  

According to this scheme, the $n$-dimensional Minkowski spacetime splits into the product of a four-dimensional subspace 
and a $(n-4)$-dimensional subspace. Any $n$-dimensional object, such as metric tensor, gamma matrices, momenta, etc., 
decomposes into a four-dimensional part (denoted by a bar) and a $(n-4)$-dimensional part (denoted by a hat), for instance
\begin{equation}
	\eta^{ab}= \bar{\eta}^{ab} + \hat{\eta}^{ab}, \quad \gamma^a = \bar{\gamma}^a + \hat{\gamma}^a, \quad p^a= \bar{p}^a + \hat{p}^a, \quad ...
\end{equation} 
Contractions of indices belonging to different subspaces vanish. 
The chiral matrix $\gamma^5$ is defined as in four dimensions by \eqref{gamma5} 
\begin{equation} 
\gamma^5= \frac{i}{4 !} \epsilon_{a b c d} \gamma^a \gamma^b \gamma^c \gamma^d = - i \bar{\gamma}^0 \bar{\gamma}^1 \bar{\gamma}^2 \bar{\gamma}^3
\label{g5}
\end{equation}
where $\epsilon_{a b c d}$ is purely four-dimensional, and anticommutes with the gamma matrices of the four-dimensional subspace, while it commutes with those belonging to the $(n-4)$-dimensional one
\begin{equation}  \begin{aligned} 
  &\left\lbrace \gamma^5, \bar{\gamma}^a\right\rbrace=0  \quad \text{for}\ a=0,1,2,3 \\
  	&\left[  \gamma^5, \hat{\gamma}^a\right] =0  \quad \text{for}\ a\geq 4.
         \end{aligned}  \end{equation}
This prescription is such to preserve the square of $\gamma^5$ to unity and the cyclicity of the trace. 
It breaks $n$-dimensional Lorentz covariance for chiral objects and, as a consequence, 
spurious noncovariant terms may appear in the calculations, 
which however are expected to be removable by finite noncovariant counterterms.

The Breitenlohner-Maison prescription is not the unique one for dealing with $\gamma^5$ in dimensional regularization,
 and a comparison between different proposals can be found in \cite{Baikov:1991qz}. However, among these, only the Breitenlohner-Maison scheme has been shown to give mathematically consistent results at arbitrary loop orders \cite{Breitenlohner:1977hr, Rufa:1990hg}.
 
 At this stage, we wish to stress an important remark.
As explained above, a prescription for $\gamma^5$ in $n$ dimensions is necessary to overcome inconsistencies in parity-odd calculations. 
However, whenever parity-even calculations are concerned, in which traces containing only an even number 
of $\gamma^5$ matrices appear, 
no such inconsistencies arise, and one can safely extend \eqref{anticommutator} to $n$ dimensions 
and use the square $(\gamma^5)^2 = \mathbb{1}$ to completely eliminate $\gamma^5$ from the traces 
\cite{Baikov:1991qz, Bonneau:1980yb, Gottlieb:1979ix}.

We conclude this appendix collecting a list of useful relations.
The metric tensor is split as $ \eta^{ab}= \bar{\eta}^{ab} + \hat{\eta}^{ab} $ with
\begin{equation}
	\eta_{ab}\eta^{ab}=n, \qquad \bar{\eta}_{ab} \bar{\eta}^{ab}=4, \qquad \hat{\eta}_{ab} \hat{\eta}^{ab}=n-4, \qquad \bar{\eta}_{ab} \hat{\eta}^{ab}=0.
\end{equation}
The last shows that contractions between indices belonging to different subspaces vanish.

Any vector decomposes as
\begin{equation}
	k^a= \bar{k}^a + \hat{k}^a
\end{equation}
and metric tensors act as projectors onto different subspaces
\begin{align}
&	k^a= \eta^{ab}k_b, \qquad k_a=\eta_{ab}k^b, \qquad \bar{k}_a=\bar{\eta}_{ab}k^b, \qquad \hat{k}_a=\hat{\eta}_{ab}k^b, \qquad  k^2 = \bar{k}^2 + \hat{k}^2,
\nonumber
\\
&  k^2=k^ak_a= \eta^{ab}k_ak_b= \eta_{ab}k^ak^b, \qquad \bar{k}^2 = \bar{k}_a \bar{k}^a= \bar{\eta}^{ab}k_a k_b=\bar{\eta}_{ab}k^a k^b,
\nonumber
\\
& \hat{k}^2 = \hat{k}_a \hat{k}^a= \hat{\eta}^{ab}k_a k_b=\hat{\eta}_{ab}k^a k^b, \qquad \bar{\eta}_{ab}\hat{k}^b=0, \qquad \hat{\eta}_{ab}\bar{k}^b=0.
\end{align}

Gamma matrices decompose as
\begin{equation}
	\gamma^a= \bar{\gamma}^a + \hat{\gamma}^a
\end{equation}
and satisfy 
\begin{align}
& \{ \gamma^a , \gamma^b \} =2 \eta^{ab}, \qquad \gamma^a \gamma_a=n, \qquad {\rm tr}\, \gamma^a =0, 
\nonumber
\\
&  \{ \gamma^a , \bar{\gamma}^b \}=\{ \bar{\gamma}^a , \bar{\gamma}^b\} = 2 \bar{\eta}^{ab}, \qquad \gamma^a \bar{\gamma}_a=\bar{\gamma}^a \bar{\gamma}_a = 4, \qquad {\rm tr}\,  \bar{\gamma}^a=0, 
\nonumber
\\
&   \{ \gamma^a , \hat{\gamma}^b \} =\{ \hat{\gamma}^a , \hat{\gamma}^b\} = 2 \hat{\eta}^{ab}, \qquad \gamma^a \hat{\gamma}_a=\hat{\gamma}^a \hat{\gamma}_a = n-4, \qquad {\rm tr}\,  \hat{\gamma}^a=0, 
\nonumber
\\
&   \{\bar{\gamma}^a , \hat{\gamma}^b \} = 0,   \qquad \bar{\gamma}^a \hat{\gamma}_a=0.
\end{align}	

The matrix $\gamma^5$ is defined as in four dimensions, see eq. \eqref{g5}. 
It anticommutes with the gamma matrices 
of the four-dimensional subspace and commutes with those of the $(n-4)$-dimensional subspace
\begin{equation}
	\left\lbrace \gamma^5, \bar{\gamma}^a \right\rbrace =0, \qquad \left[ \gamma^5, \hat{\gamma}^a\right] =0
\end{equation}
which implies 
\begin{equation}
	\left\lbrace \gamma^5, \gamma^a \right\rbrace = \left\lbrace \gamma^5, 
	\hat{\gamma}^a \right\rbrace = 2 \gamma^5  \hat{\gamma}^a, 
	\qquad \left[ \gamma^5, \gamma^a\right] = \left[ \gamma^5, \bar{\gamma}^a\right] = 2 \gamma^5 \bar{\gamma}^a.
\end{equation}

From the definition of $\gamma^5$ in \eqref{g5}, its square $(\gamma^5)^2 = \mathbb{1}$, and the definition of the 
chiral projectors \eqref{chiral projectors}, one can derive the following identities
\begin{equation} \label{id2}
 P_R \gamma^a P_L = \bar{\gamma}^a P_L = P_R\bar{\gamma}^a, \qquad P_L \gamma^a P_R = \bar{\gamma}^a P_R = P_L\bar{\gamma}^a.
\end{equation}

At last, we list the explicit expression of traces involving two, four, and six gamma matrices in $n$ even dimensions
\begin{equation} \label{trace of two gammas}
{\rm tr}\, ( \gamma^a \gamma^b ) =  2^{\frac{n}{2}} \eta^{ab}
\end{equation}

\begin{equation} \label{trace of four gammas}
{\rm tr}\, ( \gamma^a \gamma^b \gamma^c \gamma^d )= 
 2^{\frac{n}{2}} ( \eta^{ab}\eta^{cd} - \eta^{ac}\eta^{bd} + \eta^{ad} \eta^{bc} )
\end{equation}
\begin{align} \label{trace of six gammas}
 {\rm tr}\, (\gamma^d \gamma^e \gamma^a \gamma^f \gamma^b \gamma^g )  &= 
  2^{\frac{n}{2}} (   \eta^{de} \eta^{af} \eta^{bg} - \eta^{de} \eta^{ab} \eta^{fg} + \eta^{de} \eta^{ag} \eta^{fb} 
  \nonumber
  \\
&  \quad  \qquad  - \eta^{da} \eta^{ef} \eta^{bg} + \eta^{da} \eta^{eb} \eta^{fg} - \eta^{da} \eta^{eg} \eta^{fb} 
\nonumber
\\
&  \quad  \qquad   + \eta^{ea}\eta^{df} \eta^{bg}  - \eta^{ea} \eta^{db} \eta^{fg} + \eta^{ea} \eta^{dg} \eta^{fb}  
\nonumber
\\
& \quad  \qquad - \eta^{df} \eta^{eb} \eta^{ag} + \eta^{df} \eta^{ab} \eta^{eg} - \eta^{ef} \eta^{ab} \eta^{dg} 
\nonumber
\\  
& \quad \qquad + \eta^{ef} \eta^{db} \eta^{ag} - \eta^{af} \eta^{db} \eta^{eg} + \eta^{af} \eta^{eb} \eta^{dg} ) \ .
\end{align}

\section{Loop integrals and dimensional regularization} 
\label{C}
In order to combine propagator denominators in loop integrals, we have used Feynman parametric formulae
	\begin{align}
	& \frac{1}{AB} = \int_0^1 dx\, \frac{1}{\left[  x A + (1-x)B\right]^2 }  \label{Fp int 1}\\
	& \frac{1}{ABC}= 2 \int_0^1 dx \int_0^{1-x} dy\,  \frac{1}{\left[ x A + yB + (1-x-y)C\right]^3 } 
	\label{Fp int 2} 
	\end{align}
which make the denominators quadratic functions of the loop 
integration variable $l$ used in the main text. 
Then, one completes the square and shifts the integration variable to absorb linear terms in $l$. 
The denominator takes the form $(l^2 + f)^m$, where $m=2,3$ and $f$ is a function of the Feynman parameters and external momenta.
Performing integration over the loop momentum $l$, terms with odd powers of $l$ in the numerator vanish by symmetry. Symmetry allows also to replace 
	\begin{align}
	& l^a l^b \to \frac{1}{n}\eta^{ab}l^2 \\
	& l^a l^b l^c l^d \to \frac{1}{n (n+2)} \ l^4 \left( \eta^{ab} \eta^{cd} + \eta^{ac} \eta^{bd} + \eta^{ad} \eta^{bc}\right) 
	\end{align}
where $n$ is the spacetime dimension. It is most convenient to evaluate the integrals by Wick-rotating the integration variable to Euclidean space, i.e.  by replacing $l^0 \to i l^0$.

In the following table we collect $n$-dimensional integrals in Minkowski space 
\begin{equation}
\int \frac{d^n l}{(2 \pi)^n} \frac{1}{\left( l^2 + f\right) ^ m} = \frac{ i}{(4 \pi)^\frac{n}{2}} \frac{\Gamma\left( m - \frac{n}{2}\right) }{\Gamma(m)} \left( \frac{1}{f}\right) ^{m - \frac{n}{2}}
\end{equation}
\begin{equation}
\int \frac{d^n l}{(2 \pi)^n} \frac{l^2}{(l^2 + f)^m}  = \frac{i}{(4 \pi)^2} \frac{n}{2}\frac{\Gamma(m - \frac{n}{2} - 1)}{\Gamma(m)} \left( \frac{1}{f} \right)^{m-\frac{n}{2}-1} 
\end{equation}
\begin{equation}
\int \frac{d^n l}{(2 \pi)^n} \frac{l^4}{(l^2 + f)^m}= \frac{i}{(4 \pi)^{\frac{n}{2}}} \frac{n (n+2)}{4} \frac{\Gamma(m - \frac{n}{2} - 2)}{\Gamma(m)} \left( \frac{1}{f} \right)^{m-\frac{n}{2}-2} 
\end{equation}
where the overall factor $i$ comes from the Wick rotation of the integration variable. We need also the following expansion 
\begin{equation} \label{DR expansion}
\frac{\Gamma(2 - \frac{n}{2})}{f^{2 - \frac{n}{2}}} \overset{n \to 4}{=} \frac{2}{4-n} - \log f - \gamma + \log 4 \pi + O(4-n) 
\end{equation}
where $\gamma$ is the Euler-Mascheroni constant. Thanks to this table we can easily evaluate integrals appearing in the calculations. 

Splitting the loop momentum $l= \bar{l} + s$ we can evaluate the integrals
\begin{equation} 
\int \frac{d^4 \bar{l}}{(2 \pi)^4} \int \frac{d^{n-4} s}{(2 \pi)^{n-4}} \frac{s^2}{(\bar{l}^2 + s^2 + f)^3}
\end{equation}
and
\begin{equation}
\int \frac{d^4 \bar{l}}{(2 \pi)^4} \int \frac{d^{n-4} s}{(2 \pi)^{n-4}} \frac{s^4}{(\bar{l}^2 + s^2 + f)^3} \  .
\end{equation}
Let us first perform integration over $s$ and define $t= \bar{l}^2 + f$
\begin{equation}
\int \frac{d^{n-4} s}{(2 \pi)^{n-4}} \frac{s^2}{(s^2 + t)^3} = \frac{n-4}{2 (4 \pi)^\frac{n-4}{2}} \frac{\Gamma(4 -\frac{n}{2})}{\Gamma(3)} \left( \frac{1}{t}\right)^{4 - \frac{n}{2}}
\end{equation}
then, integrating over $\bar{l}$
\begin{equation}
\int \frac{d^4 \bar{l}}{(2 \pi)^4} \frac{1}{( \bar{l}^2 + f)^{4 -\frac{n}{2}}}= \frac{i}{(4 \pi)^2} \frac{\Gamma(2 - \frac{n}{2})}{\Gamma(4 - \frac{n}{2})}\left( \frac{1}{f}\right)^{2 - \frac{n}{2}} \ .
\end{equation}
Putting everything together and using the expansion \eqref{DR expansion} we get the finite result
\begin{equation} \label{DR int 4}
\int \frac{d^4 \bar{l}}{(2 \pi)^4} \int \frac{d^{n-4} s}{(2 \pi)^{n-4}} \frac{s^2}{(\bar{l}^2 + s^2 + f)^3}= - \frac{i}{32 \pi^2}
\;.
\end{equation}

Following similar steps one has
\begin{equation}
\int \frac{d^{n-4} s}{(2 \pi)^{n-4}} \frac{s^4}{( s^2 + t)^3} = \frac{(n-4)(n-2)}{(4 \pi)^{\frac{n-4}{2}}4} \frac{\Gamma(3 - \frac{n}{2})}{\Gamma(3)} \left( \frac{1}{t}\right)^{3 - \frac{n}{2}} 
\end{equation}
from the integration over $s$, and
\begin{equation}
\int \frac{d^4 \bar{l}}{(2 \pi)^4} \frac{1}{(\bar{l}^2 + f)^{3 - \frac{n}{2}}} = \frac{i}{( 4 \pi)^2} \frac{\Gamma(1 -\frac{n}{2})}{\Gamma(3 -\frac{n}{2})} \left( \frac{1}{f}\right)^{1 -\frac{n}{2}} 
\end{equation}
from integrating over $\bar{l}$.  Putting everything together we obtain the finite result
\begin{equation} \label{DR int 5}
\int \frac{d^4 \bar{l}}{(2 \pi)^4} \int \frac{d^{n-4} s}{(2 \pi)^{n-4}} \frac{s^4}{(\bar{l}^2 + s^2 + f)^3}=  \frac{i}{32 \pi^2} f \ .
\end{equation}
We notice that the factor $(n-4)$ arising from the integration over $s$ kills the singularity in the expansion \eqref{DR expansion} and cancels all other terms, once the limit $n \to 4$ is taken, leading to a finite result.

For completeness, we collect here the relevant integrals employed during the calculations 
derived from the above table
\begin{align} \label{DR int 1}
\left( \frac{2}{n} - 1\right) \int \frac{d^n l}{(2 \pi)^n} \frac{l^2}{(l^2 + f)^2} & = \left( \frac{2}{n} - 1\right) \frac{i}{(4 \pi)^{\frac{n}{2}}} \frac{n}{2} \frac{\Gamma(1 - \frac{n}{2})}{\Gamma(2)} \left( \frac{1}{f}\right)^{1- \frac{n}{2}} 
\nonumber
 \\
& = \frac{i}{(4 \pi)^{\frac{n}{2}}}  \left( 1 - \frac{n}{2} \right) \frac{\Gamma(1 - \frac{n}{2})}{f^{1- \frac{n}{2}}}  
\nonumber
\\
& =  \frac{i}{(4 \pi)^{\frac{n}{2}}} \frac{\Gamma(2 - \frac{n}{2})}{f^{2- \frac{n}{2}}} f  
\nonumber
\\
& \overset{n \to 4}{=} \frac{i}{16 \pi^2} f \left( \frac{2}{4-n} - \log f - \gamma + \log 4\pi + O(4-n)\right)
\end{align}

\begin{align} \label{DR int 2}
\int \frac{d^n l}{(2 \pi)^n} \frac{1}{(l^2 + f)^2} & = \frac{i}{(4 \pi)^2} \frac{\Gamma(2 - \frac{n}{2})}{\Gamma(2)} \left( \frac{1}{f} \right)^{2-\frac{n}{2}}  
\nonumber
\\
& \overset{n\to 4}{=} \frac{i}{16 \pi^2} \left( \frac{2}{4-n} - \log f - \gamma + \log 4\pi + O(4-n)\right)
\end{align}

\begin{align} \label{DR int 3}
\int \frac{d^n l}{(2 \pi)^n} \frac{l^2}{(l^2 + f)^3} & = \frac{i}{(4 \pi)^2} \frac{n}{2}\frac{\Gamma(2 - \frac{n}{2})}{\Gamma(3)} \left( \frac{1}{f} \right)^{2-\frac{n}{2}}  
\nonumber
\\
& \overset{n\to 4}{=} \frac{i}{16 \pi^2} \left( \frac{2}{4-n} - \log f - \gamma + \log 4\pi + O(4-n)\right) \ .
\end{align}

\section{Stress tensor conservation} 
\label{D}
Let us compute
\begin{subequations}
	\begin{align}
	- i k_c \mathcal{T}^{cdab} = & - \frac{1}{4} \int \frac{d^n l}{(2 \pi)^n} {\rm tr} \Big(\gamma^d \slashed{l} \gamma^a   (\slashed{l} - \slashed{p}) \gamma^b (\slashed{l} - \slashed{k})\Big)   
	 \left( \frac{1}{(l-p)^2 (l-k)^2} - \frac{1}{l^2 (l-p)^2}\right)   \label{one}  \\
	&	- \frac{1}{4} \int \frac{d^n l}{(2 \pi)^n} \left( 2l -k\right)^d \left(  \frac{{\rm tr}\left\lbrace \gamma^a (\slashed{l} - \slashed{p}) \gamma^b (\slashed{l} - \slashed{k})\right\rbrace }{(l-p)^2 (l-k)^2} - \frac{{\rm tr}\left\lbrace \slashed{l}\gamma^a (\slashed{l} - \slashed{p})\gamma^b\right\rbrace }{l^2 (l-p)^2}\right) \label{two} \\
	& + \frac{1}{4} \ k^b  \int \frac{d^n l}{(2 \pi)^n} \frac{{\rm tr}\left\lbrace \gamma^d \slashed{l}\gamma^a (\slashed{l} - \slashed{p})\right\rbrace }{l^2 (l-p)^2} \label{three} \\
	& + \frac{1}{4} \ \eta^{bd} \int \frac{d^n l}{(2 \pi)^n}  \frac{{\rm tr} \left\lbrace \slashed{q} \slashed{l}\gamma^a (\slashed{l} - \slashed{p})\right\rbrace }{l^2 (l-p)^2} \label{four} \\
	& + \frac{1}{4} \ k^a\int \frac{d^n l}{(2 \pi)^n}  \frac{{\rm tr}\left\lbrace \gamma^d \slashed{l}\gamma^b (\slashed{l} - \slashed{q})\right\rbrace }{l^2 (l-q)^2} \label{five} \\
	& + \frac{1}{4} \ \eta^{ad}\int \frac{d^n l}{(2 \pi)^n}  \frac{{\rm tr}\left\lbrace \slashed{p} \slashed{l}\gamma^b (\slashed{l} - \slashed{q})\right\rbrace }{l^2 (l-q)^2} \label{six} 
	\end{align}
\end{subequations}
The terms \eqref{three} and \eqref{five} have the same structure of the integral appearing in the 
vacuum polarization (photon self-energy) and yield
\begin{subequations}
	\begin{align}
	& \frac{i}{8 \pi^2}(p + q)^b \left( p^a p^d - \eta^{ad}p^2\right) \int_{0}^{1} dx \ x(x-1) \left( \frac{2}{4-n} - \log f - \gamma + \log 4\pi + O(4-n)\right)  \label{f1}\\
	& + \frac{i}{8 \pi^2}(p + q)^a \left( q^b q^d - \eta^{bd}q^2\right) \int_{0}^{1} dx \ x(x-1) \left( \frac{2}{4-n} - \log g - \gamma + \log 4\pi + O(4-n)\right) \label{g1} 
	\end{align}
\end{subequations}
where $f\equiv f(x,p)= p^2 x(1-x)$, $g\equiv g(q,x)= q^2x(1-x)$, $\gamma$ is the Euler-Mascheroni constant and we have rewritten $k = p + q$ due to momentum conservation.

Let us consider \eqref{four}. Using Feynman parametric formula \eqref{Fp int 1} it becomes
\begin{equation}
\frac{1}{4} \ \eta^{bd} \int_0^1 dx \int \frac{d^n l}{(2 \pi)^n}  \frac{{\rm tr}\left\lbrace \slashed{q} (\slashed{l} + \slashed{p} x)\gamma^a (\slashed{l} + \slashed{p}(x-1))\right\rbrace }{ (l^2 + f)^2}
\end{equation}
where $f\equiv f(x,p)= p^2 x(1-x)$ and integration variable shifted by $l \to l + px$. Then, we evaluate the trace 
\begin{equation}
 {\rm tr}\left\lbrace \gamma^e \gamma^f \gamma^a \gamma^g\right\rbrace  q_e (l+px)_f (l+p(x-1))_g = {\rm tr}\left\lbrace \gamma^e \gamma^f \gamma^a \gamma^g\right\rbrace \Big(q_e l_f l_g + q_e p_f p_g x(x-1)\Big)
\end{equation}
where  linear terms in $l$ have been neglected because they vanish by symmetric integration. After replacing $l_f l_g \to \frac{1}{n}\eta_{fg}l^2$, using \eqref{trace of four gammas}, \eqref{DR int 1} and \eqref{DR int 2} we obtain
\begin{equation} \label{f5}
\begin{split}
& - \frac{i}{16 \pi^2} \eta^{bd} \Big(2 q^a p^2 -2pq p^a\Big) \int_{0}^1 dx \ x(x-1)  \left( \frac{2}{4-n} - \log f - \gamma + \log 4\pi + O(4-n)\right).
\end{split}
\end{equation}

By an analogous reasoning, we can compute \eqref{six} which yields
\begin{equation}
\begin{split} \label{g_5}
& - \frac{i}{16 \pi^2} \eta^{ad} \Big(2 p^b q^2 -2pq q^b\Big) \int_{0}^1 dx \  x(x-1)  \left( \frac{2}{4-n} - \log g - \gamma + \log 4\pi + O(4-n)\right)
\end{split}
\end{equation}
where $g\equiv g(q,x)= q^2x(1-x)$.

Let us now consider the first integral of \eqref{two}, after using the Feynman parametric formula and shifting the integration variable $l \to l+p+ qx$, this becomes
\begin{equation}
-\frac{1}{4} \int_{0}^{1} dx \int \frac{d^n l}{(2 \pi)^n} \Big( 2l + p + q(2x-1)\Big)^d \ \frac{{\rm tr}\left\lbrace \gamma^a (\slashed{l} + \slashed{q}x) \gamma^b (\slashed{l} + \slashed{q}(x-1))\right\rbrace }{(l^2 + g)^2} 
\end{equation}
where $g\equiv g(q,x)= q^2 x(1-x)$. Terms proportional to $(2x-1)$ vanish by integrating over $x$. 
The integral proportional to $p^d$ has the same structure of the photon self-energy and yields
\begin{equation} \label{g2}
- \frac{i}{8 \pi^2}p^d \left( q^a q^b - \eta^{ab}q^2\right) \int_{0}^{1} dx \ x(x-1)\left( \frac{2}{4-n} - \log g - \gamma + \log 4\pi + O(4-n)\right).
\end{equation}
Now, we evaluate
\begin{equation}
-\frac{1}{2} \int_{0}^{1} dx \int \frac{d^n l}{(2 \pi)^n}  l^d  \frac{{\rm tr}\left\lbrace \gamma^a (\slashed{l} + \slashed{q}x) \gamma^b (\slashed{l} + \slashed{q}(x-1))\right\rbrace }{(l^2 + g)^2} \ .
\end{equation}
Let us compute the trace using \eqref{trace of four gammas} 
\begin{equation}
\begin{split}
& {\rm tr}\left\lbrace \gamma^a \gamma^c \gamma^b \gamma^e\right\rbrace \left( l+qx\right)_c \left( l + q(x-1) \right)_e = \\
& = 2^{\frac{n}{2}} \left(  \left(l + qx \right)^a \left( l +q (x-1) \right)^b +  \left(l + qx \right)^b \left( l +q (x-1) \right)^a - \eta^{ab} \left( l+qx\right)  \left( l + q(x-1)\right) \right)  \ .
\end{split}
\end{equation}
Since the integral is non zero only if even powers of $l$ appear in the numerator, we keep only terms of this trace with one $l$, and obtain
\begin{equation}
-  \int_0^1 dx (2x-1) \int \frac{d^n l}{(2 \pi)^n}  \frac{l^d l^a q^b + l^d l^b q^a - \eta^{ab}l^d l q}{(l^2 + g)^2} =0
\end{equation}
which vanishes by integration over $x$.

Let us now focus on the second integral of \eqref{two} which can be rewritten as
\begin{equation}
\frac{1}{4} \int_0^1 dx \int \frac{d^n l}{(2 \pi)^n} \left( 2l +p(2x-1) -q\right)^d  \frac{{\rm tr}\left\lbrace (\slashed{l} + \slashed{p}x) \gamma^a (\slashed{l} + \slashed{p}(x-1)) \gamma^b\right\rbrace }{(l^2 + f)^2} 
\end{equation}
with $f \equiv f(x,p)= p^2 x(1-x)$. By following a similar reasoning as before, the unique non zero term is
\begin{equation} \label{f2}
-\frac{i}{8 \pi^2}q^d \left( p^a p^b - \eta^{ab}p^2\right) \int_{0}^{1} dx\  x(x-1)\left( \frac{2}{4-n} - \log f - \gamma + \log 4\pi + O(4-n)\right).
\end{equation}
Let us now evaluate \eqref{one} and start from the first term
\begin{equation}
-\frac{1}{4} \int \frac{d^n l}{(2 \pi)^n}  \frac{{\rm tr} \left\lbrace \gamma^d \slashed{l} \gamma^a (\slashed{l} - \slashed{p}) \gamma^b (\slashed{l} - \slashed{k})\right\rbrace }{(l-p)^2 (l-k)^2} \ .
\end{equation}
Introducing Feynman parameter and shifting $l \to l + p + qx$, this becomes
\begin{equation}
-\frac{1}{4} \int_0^1 dx \int \frac{d^n l}{(2 \pi)^n} \frac{{\rm tr} \left\lbrace \gamma^d (\slashed{l} + \slashed{p} + \slashed{q}x) \gamma^a (\slashed{l} + \slashed{q}x) \gamma^b (\slashed{l} + \slashed{q}(x-1))\right\rbrace }{(l^2+g)^2}
\end{equation}
where $g \equiv g(q,x)= q^2 x(1-x)$. In evaluating the trace we neglect terms containing an odd number of $l$ because they vanish by symmetric integration. Thus, one has
\begin{equation}
{\rm tr} \left\lbrace \gamma^d \gamma^e \gamma^a \gamma^f \gamma^b \gamma^g\right\rbrace  \Big( l_e l_f q_g(x-1) + l_el_g q_fx + l_f l_g (p + qx)_e + (p + qx)_e q_f x q_g(x-1)\Big) \ .
\end{equation}
By symmetric integration we can replace $l_a l_b \to \frac{1}{n}\eta_{ab}l^2$, use \eqref{trace of six gammas} and compute
\begin{align}
& {\rm tr} \left\lbrace \gamma^d \gamma^e \gamma^a \gamma^f \gamma^b \gamma^g\right\rbrace l_e l_f  q_g(x-1) 
\nonumber
\\ 
& = \frac{1}{n} l^2 {\rm tr} \left\lbrace \gamma^d \gamma^e \gamma^a \gamma^f \gamma^b \gamma^g \right\rbrace  \eta_{ef} q_g(x-1) 
\nonumber
\\
& = \frac{1}{n} l^2 2^{\frac{n}{2}} \left(  \eta^{ab}(2-n) q^d(x-1) + \eta^{db}(n-2)  q^a(x-1) + \eta^{da}(2-n) q^b(x-1) \right) 
\nonumber
 \\
& = \left( \frac{2}{n} - 1 \right) 2^{\frac{n}{2}} l^2 \left(  \eta^{ab} q^d(x-1) - \eta^{db} q^a(x-1) + \eta^{da}q^b(x-1) \right) \ ,
\end{align}

\begin{align}
& {\rm tr} \left\lbrace \gamma^d \gamma^e \gamma^a \gamma^f \gamma^b \gamma^g \right\rbrace  l_f l_g (p + q x)_e =  \frac{1}{n} l^2 {\rm tr} \left\lbrace \gamma^d \gamma^e \gamma^a \gamma^f \gamma^b \gamma^g\right\rbrace  \eta_{fg} (p + q x)_e  
\nonumber
\\
& = \left( \frac{2}{n} - 1\right) 2^{\frac{n}{2}} l^2 \left(  \eta^{ab} (p + q x)^d - \eta^{da}(p + q x)^b + \eta^{db} (p + q x)^a \right) \ ,
\end{align}

\begin{align}
& {\rm tr} \left\lbrace \gamma^d \gamma^e \gamma^a \gamma^f \gamma^b \gamma^g \right\rbrace  l_e l_g q_f x = 
\frac{1}{n} l^2 {\rm tr} \left\lbrace \gamma^d \gamma^e \gamma^a \gamma^f \gamma^b \gamma^g\right\rbrace  \eta_{eg} q_f x 
\nonumber
\\
& =\left( \frac{2}{n} - 1\right) 2^{\frac{n}{2}} l^2 \left(  \eta^{db} q^a x + \eta^{da} q^b x- \eta^{ab} q^d x \right) \ .
\end{align}
Putting everything together we find
\begin{equation}
\left( \frac{2}{n} - 1\right) 2^{\frac{n}{2}} l^2 \left(  \eta^{ab} \left( q^d (x-1) + p^d\right) + \eta^{da}\left( - p^b + q^b(x-1)\right) + \eta^{db} \left( q^a(x+1) + p^a\right) \right) \ .
\end{equation}
Integrating over $l$ using \eqref{DR int 1}, one obtains
\begin{align} \label{g3}
& \frac{i}{16 \pi^2} \int_0^1 dx \Big(\eta^{ab} \left( q^d (x-1) + p^d\right) + \eta^{da}\left( - p^b + q^b(x-1)\right) + 
\nonumber
\\
& + \eta^{db} \left( q^a(x+1) + p^a\right)  \Big) 
g \left( \frac{2}{4-n} - \log g - \gamma + \log 4\pi + O(4-n)\right) \ .
\end{align}
From the term with no $l$ we obtain
\begin{align}
& {\rm tr} \left\lbrace \gamma^d \gamma^e \gamma^a \gamma^f \gamma^b \gamma^g\right\rbrace  \left( p +qx \right)_e q_f q_g x(x-1) = \\
& = 2^{\frac{n}{2}} x(x-1) \Big( \left( p + qx\right)^d \left( 2 q^a q^b - \eta^{ab}q^2\right)  +
\nonumber
 \\
& \quad + \left( p + qx\right)^a \left( 2 q^d q^b - \eta^{db}q^2\right) + \eta^{ad} \left( q^2( p + qx)^b - 2q^b (p+qx) q \right) \Big)
\end{align}
and after integrating over $l$ using \eqref{DR int 2}
\begin{align} \label{g4}
& - \frac{i}{16 \pi^2} \int_{0}^1 dx \ x(x-1) \times 
\nonumber
\\
& \quad \times \Big( \left( p + qx\right)^d \left( 2 q^a q^b - \eta^{ab}q^2\right)  +   \left( p + qx\right)^a \left( 2 q^d q^b - \eta^{db}q^2\right) +
\nonumber
 \\
& \qquad + \eta^{ad} \left( q^2( p + qx)^b - 2q^b (p+qx)  q \right) \Big) \times 
\nonumber
\\
& \quad \qquad  \times \left( \frac{2}{4-n} - \log g - \gamma + \log 4\pi + O(4-n)\right) \ .
\end{align}
By adding \eqref{g1}, \eqref{g_5}, \eqref{g2}, \eqref{g3} and \eqref{g4} we obtain
\begin{align}
& - \frac{i}{16 \pi^2} \int_{0}^1 dx \ x(x-1) \times
\nonumber
 \\
& \times \Big(  4p^d \left( q^a q^b - \eta^{ab}q^2\right) + 4 \eta^{ad} \left( q^2 p^b - q^b p  q\right) + 
\nonumber
\\
& \quad (2x-1) \left( -\eta^{ad} q^b q^2 - \eta^{ab}q^d q^2 + 2 q^a q^b q^d + \eta^{bd}q^a q^2 \right) \Big) \times 
\nonumber
\\
& \qquad  \times \left( \frac{2}{4-n} - \log g - \gamma + \log 4\pi + O(4-n)\right) 
\end{align}
and after integrating over $x$ 
\begin{equation} \label{c_1}
\frac{i}{24 \pi^2}\Big(  p^d \left( q^a q^b - \eta^{ab}q^2\right) +  \eta^{ad} \left( q^2 p^b - q^b p  q\right) \Big)  \left( \frac{2}{4-n} + \frac{5}{3} - \log q^2 - \gamma + \log 4\pi + O(4-n)\right) \ .
\end{equation}

Let us now consider the second term of \eqref{one}
\begin{equation}
\frac{1}{4} \int \frac{d^n l}{(2 \pi)^n} \frac{{\rm tr} \left\lbrace \gamma^d \slashed{l} \gamma^a (\slashed{l} - \slashed{p}) \gamma^b (\slashed{l} - \slashed{k})\right\rbrace }{l^2 (l-p)^2}  \ .
\end{equation} 
Making use of Feynman parametric formula for rewriting the denominator and shifting the integration variable $l \to l + px$, this becomes
\begin{equation}
\frac{1}{4} \int_0^1 dx \int \frac{d^n l}{(2 \pi)^n}  \frac{ {\rm tr} \left\lbrace \gamma^d (\slashed{l} + \slashed{p}x) \gamma^a (\slashed{l} + \slashed{p}(x-1)) \gamma^b (\slashed{l} - \slashed{q} + \slashed{p}(x-1))\right\rbrace }{ (l^2+f)^2} \ .
\end{equation}
As before, we keep only terms containing an even number of $l$, and they are
\begin{equation}
{\rm tr} \left\lbrace \gamma^d \gamma^e \gamma^a \gamma^f \gamma^b \gamma^g\right\rbrace  \Big( l_e l_f (-q + p(x-1))_g + l_el_g p_f(x-1) + l_f l_g p_e x + p_e x p_f (x-1)(-q + p(x-1))_g\Big) \ .
\end{equation}
We compute
\begin{align}
& {\rm tr} \left\lbrace \gamma^d \gamma^e \gamma^a \gamma^f \gamma^b \gamma^g \right\rbrace  l_e l_f (-q + p(x-1))_g = 
\nonumber
\\
& = \Big(\frac{2}{n} - 1\Big) 2^{\frac{n}{2}} l^2 \left(  \eta^{ab}(p(x-1) - q)^d - \eta^{db}(p(x-1) - q)^a + \eta^{da}(p(x-1) - q)^b \right) \ ,
\end{align}

\begin{equation}
 {\rm tr} \left\lbrace \gamma^d \gamma^e \gamma^a \gamma^f \gamma^b \gamma^g \right\rbrace  l_el_g p_f(x-1)  = \Big(\frac{2}{n} - 1\Big) 2^{\frac{n}{2}} l^2 \left(  \eta^{db}p^a (x-1) + \eta^{da}p^b(x-1) - \eta^{ab}p^d(x-1) \right) \ ,
\end{equation}

\begin{equation}
 {\rm tr} \left\lbrace \gamma^d \gamma^e \gamma^a \gamma^f \gamma^b \gamma^g\right\rbrace  l_f l_gp_e x= \Big(\frac{2}{n} - 1\Big) 2^{\frac{n}{2}} l^2  \left(  \eta^{ab}p^d x - \eta^{da}p^b x + \eta^{db}p^a x \right) \ .
\end{equation}
After putting everything together and integrating over $l$, one has
\begin{align} \label{f3}
& - \frac{i}{16 \pi^2} \int_0^1 dx \Big( \eta^{ab}\left( p^d x - q^d\right)  + \eta^{da}\left( p^b (x-2) - q^b\right) + \eta^{db}\left( p^a x + q^a\right) \Big) f  \ \times 
\nonumber
\\
& \qquad \qquad \qquad \times \left( \frac{2}{4-n} - \log f - \gamma + \log 4\pi + O(4-n)\right) \ .
\end{align}
From the term with no $l$ one obtains
\begin{align}
& {\rm tr} \left\lbrace \gamma^d \gamma^e \gamma^a \gamma^f \gamma^b \gamma^g\right\rbrace  p_e x p_f (x-1)(-q + p(x-1))_g = 
\nonumber
\\ 
& =2^{\frac{n}{2}} x(x-1) \Big(  \left( -q + p (x-1)\right)^b \left(2 p^a p^d - \eta^{ad}p^2 \right) +
\nonumber
\\
& + \left( -q + p (x-1)\right)^d \left(2 p^a p^b - \eta^{ab}p^2 \right)  - \eta^{bd}p^a p^2 (x-1) + \eta^{bd}\left( 2 p^a p  q - q^a p^2\right) \Big)
\end{align}
and integrating over $l$
\begin{align} \label{f4}
& \frac{i}{16 \pi^2} \int_0^1 dx \  x(x-1) \times 
\nonumber
\\
& \quad \times \Big(  \left( -q + p (x-1)\right)^b \left(2 p^a p^d - \eta^{ad}p^2 \right) + \left( -q + p (x-1)\right)^d \left(2 p^a p^b - \eta^{ab}p^2 \right) + 
\nonumber
\\
& \quad \qquad - \eta^{bd}p^a p^2 (x-1) + \eta^{bd}\left( 2 p^a p  q - q^a p^2\right) \Big) \times 
\nonumber
\\
& \qquad \qquad \times \left( \frac{2}{4-n} - \log f - \gamma + \log 4\pi + O(4-n)\right).
\end{align}
Adding \eqref{f1}, \eqref{f5}, \eqref{f2}, \eqref{f3} and \eqref{f4}, we obtain
\begin{align}
& \frac{i}{16 \pi^2} \int_{0}^1 dx \  x(x-1) \times 
\nonumber
\\
& \times \Big( -4 q^d \left(p^a p^b - \eta^{ab}p^2\right)  - 4 \eta^{bd}\left(q^a p^2 -  p^a p q  \right) + 
\nonumber
\\
& \quad + (2x-1) \left( 2 p^a p^b p^d - \eta^{ab}p^2 p^d - \eta^{bd}p^a p^2 - p^b \eta^{ad}p^2\right)  \Big) \times 
\nonumber
\\
& \qquad \qquad \times \left( \frac{2}{4-n} - \log f - \gamma + \log 4\pi + O(4-n)\right)
\end{align}
and integrating over $x$
\begin{equation} \label{c_2}
\frac{i}{24 \pi^2} \Big(  q^d \left( p^a p^b - \eta^{ab}p^2\right)  +  \eta^{bd}\left(  q^a p^2 -  p^a p  q \right)  \Big) \left( \frac{2}{4-n} + \frac{5}{3}- \log p^2 - \gamma + \log 4\pi + O(4-n)\right) \ .
\end{equation}

The final result in momentum space is given by the sum of \eqref{c_1} and \eqref{c_2} which leads to \eqref{cms}.


\end{document}